\documentclass[preprint,5p,twocolumn]{elsarticle}

\usepackage[numbers]{natbib}
\usepackage{amsmath,amssymb,amsfonts}
\usepackage{graphicx}
\usepackage{textcomp}
\usepackage{xcolor}

\usepackage{subcaption}  
\usepackage[utf8]{inputenc}  
\usepackage{tabularx}  
\usepackage{makecell} 
\usepackage{multirow}  
\usepackage{booktabs}  
\usepackage[acronym]{glossaries}  
\usepackage{url}  
\usepackage{algorithm}  
\usepackage{algpseudocode}  
\usepackage{siunitx}
\usepackage{adjustbox}

\newcommand{\ForAllDo}[2]{\algorithmicforall\ #1\ \algorithmicdo\ #2}

\newacronym{AI}{AI}{Artificial Intelligence}
\newacronym{NLP}{NLP}{Natural Language Processing}
\newacronym{LLM}{LLM}{Large Language Model}
\newacronym{ICL}{ICL}{in-context learning}
\newacronym{SLO}{SLO}{Service Level Objective}
\newacronym{LoRA}{LoRA}{Low-Rank Adaptation}
\newacronym{QoS}{QoS}{Quality of Service}
\newacronym{TPOT}{TPOT}{Time Per Output Token}
\newacronym{ITL}{ITL}{Inter-Token Latency}
\newacronym{TTFT}{TTFT}{Time To First Token}
\newacronym{FCFS}{FCFS}{First Come First Served}
\newacronym{LRU}{LRU}{Least Recently Used}
\newacronym{SMAPE}{SMAPE}{Symmetric Mean Absolute Percentage Error}
\newacronym{ILP}{ILP}{Integer Linear Programming}
\newacronym{ML}{ML}{machine learning}

\journal{Future Generation Computer Systems}

\begin{document}

\begin{frontmatter}

\title{Data-Driven Optimization of GPU efficiency for Distributed LLM–Adapter Serving}

\author[1,2]{Ferran Agulló\corref{cor1}\fnref{fn1}}
\ead{ferran.agullo@bsc.es}
\author[1,2]{Joan Oliveras}
\ead{joan.oliveras@bsc.es}
\author[3]{Chen Wang}
\ead{chen.wang1@ibm.com}
\author[1]{Alberto Gutierrez-Torre}
\ead{alberto.gutierrez@bsc.es}
\author[3]{Olivier Tardieu}
\ead{tardieu@ibm.com}
\author[3]{Alaa Youssef}
\ead{asyousse@ibm.com}
\author[1,2]{Jordi Torres}
\ead{jordi.torres@bsc.es}
\author[1,2]{Josep Ll. Berral}
\ead{josep.ll.berral@upc.edu}
\cortext[cor1]{Corresponding author}
\affiliation[1]{organization={Barcelona Supercomputing Center (BSC)}, city={Barcelona}, country={Spain}}
\affiliation[2]{organization={Universitat Politècnica de Catalunya - BarcelonaTech (UPC)}, city={Barcelona}, country={Spain}}
\affiliation[3]{organization={IBM Research}, city={New York}, country={USA}}
\fntext[fn1]{Code will be available at: \url{https://github.com/FerranAgulloLopez/DistributedEfficientAdapterLLMServing}}

\begin{abstract}
Large Language Model (LLM) adapters enable low-cost model specialization, but introduce complex caching and scheduling challenges in distributed serving systems where hundreds of adapters must be hosted concurrently. While prior work has largely focused on latency and throughput optimization, minimizing GPU resource requirements through near-peak utilization remains largely underexplored. This paper presents a data-driven pipeline that, for a given workload, computes an adapter placement that serves the workload with the minimum number of GPUs while avoiding request starvation and GPU memory errors. To that end, the approach identifies the maximum feasible throughput attainable on each GPU by leveraging accurate performance predictions learned from real serving behavior. The proposed pipeline integrates three components: (i) a Digital Twin (DT) tailored to LLM-adapter serving, (ii) a distilled machine learning (ML) model trained on DT-generated data, and (iii) a greedy placement algorithm that exploits ML-based performance estimates to maximize GPU efficiency. The DT emulates real system dynamics with high fidelity, achieving below 5\% throughput estimation error while executing up to 90× faster than full LLM benchmarking across both predictable and unpredictable workloads. The learned ML models further accelerate performance estimation with marginal accuracy degradation, enabling scalable optimization. Experimental results demonstrate that the pipeline substantially improves GPU efficiency, reducing the number of GPUs required to sustain target workloads by 60\% on average across the evaluated scenarios. Beyond GPU efficiency, the pipeline can be adapted to alternative objectives, such as latency minimization, highlighting its versatility for future large-scale LLM serving infrastructures.
\end{abstract}


\end{frontmatter}

\section{Introduction}
\errorcontextlines=999
With the rapid advancement and widespread adoption of \Glspl{LLM}, the demand for \Gls{LLM}-adapters has grown significantly. While \Glspl{LLM} are large-scale models trained to achieve strong performance across diverse language tasks, adapters specialize this general knowledge to concrete applications through lightweight parameter additions~\cite{hu2022lora, houlsby2019parameter, liu2022few, guo2020parameter}. The use of adapters is far quicker than training-from-scratch or fine-tuning a new model, which is a tedious and long process that requires meticulous data curation and high computation capabilities. Moreover, adapters achieve comparable performance to other equivalent methods such as in-context learning~\cite{brown2020language} or prompt tuning~\cite{li2021prefix, qin2021learning}.

As adapters are increasingly used to specialize and deploy \Glspl{LLM} across diverse tasks~\cite{Wang2025}, efficiently managing their inference execution has become a key systems concern~\cite{wu2024dlora, li2025toppings, iliakopoulou2024chameleon}. Although adapters were originally designed to be merged into the backbone model to avoid additional computation and preserve latency~\cite{hu2022lora}, contemporary serving systems typically keep them unmerged. This design enables the serving of multiple task-specific adapters on top of a shared backbone without replicating its large parameter footprint in GPU memory. Such an approach is particularly well suited for multi-tenant environments, where a system must serve multiple users, each requiring a distinct model specialization~\cite{10.1145/3711875.3729141}. Since adapters are compact, a single GPU can serve hundreds or even thousands of different specializations of the same \Gls{LLM}~\cite{sheng2024slora, bruel2024compress}. This consolidation increases per-GPU throughput by aggregating requests from the hosted adapters whose individual arrival rates are often too low to saturate a GPU.

However, excessive adapter packing can cross a critical threshold where \emph{request starvation} emerges. In this regime, the GPU lacks sufficient memory for intermediate request states (i.e. KV cache), so requests accumulate faster than they are processed and latency increases steadily. Determining an allocation that maximizes per-GPU throughput, and thus GPU utilization, without triggering starvation is non-trivial. This optimal point, denoted as \(Max_{pack}\), depends on the interplay between the sizes and request arrival rates of the adapters, which vary across deployments and over time. Adapter size constrains the GPU memory available for intermediate states and increases computation time, while arrival rate governs the load induced by each adapter.

To mitigate the GPU footprint of adapter weights, serving systems often employ dynamic mechanisms that swap adapters between CPU and GPU memory based on utilization. Adapters resident in GPU memory can execute requests in parallel with other resident adapters~\cite{chen2024punica}, while swapping enables interleaved execution among adapters that do not simultaneously coexist on the GPU. Some frameworks, such as vLLM~\cite{kwon2023efficient}, additionally enforce a static upper bound on the number of loaded adapters, \(A_{max}\), and preallocate the corresponding GPU memory at initialization. This choice directly shapes \(Max_{pack}\): increasing \(A_{max}\) reduces memory available for request processing, while decreasing it limits achievable parallelism. Improper tuning of \(A_{max}\) can therefore induce starvation and, in extreme cases, GPU memory errors.

Building on these observations, this work addresses the following problem: \textbf{given an expected future workload composed of adapters with heterogeneous sizes and request arrival rates, determine a GPU allocation strategy that maximizes per-GPU throughput by achieving \(Max_{pack}\), so that the workload is served using the minimum number of GPUs without incurring starvation or memory errors}. The solution must also derive the corresponding \(A_{max}\) configuration for each GPU. By \textit{expected future workload}, we focus on \emph{predictable} workloads that can be estimated in advance, such as long-term patterns present in production traces that exhibit periodicity~\cite{wu2024dlora}. This formulation enables distributed systems to provision the minimal GPU capacity required for a workload in advance, improving hardware efficiency. We refer to this optimization challenge as the \textbf{adapter caching problem}, whose output specifies adapter placement across GPUs together with per-GPU \(A_{max}\) values.

While prior work has explored the optimization of \Gls{LLM}-adapter serving via kernel-level enhancements~\cite{chen2024punica}, memory management techniques~\cite{sheng2024slora}, and scheduling strategies~\cite{iliakopoulou2024chameleon}, the adapter caching problem remains largely underexplored. The most closely related approaches, such as dLoRA~\cite{wu2024dlora} and LoRAServe~\cite{jaiswal2025serving}, propose proactive adapter placement strategies based on estimated long-term workload patterns, aiming to reduce latency or improve throughput by fully leveraging available hardware resources. In contrast, our objective is to maximize hardware efficiency by minimizing the number of required GPUs through near-peak utilization of a subset of the devices, while leaving the remaining GPUs available for alternative workloads or reduced energy consumption.

To this end, we propose a data-driven pipeline comprising three phases: (i) a \emph{Digital Twin} (DT) that emulates an online \Gls{LLM}-adapter serving system; (ii) a machine learning (ML) phase that employs data generated by the DT; and (iii) a greedy algorithm that computes the final adapter placement to solve the caching problem. Unlike prior approaches, which rely on heuristics or partial knowledge of serving behavior, our method bases allocation decisions on complete serving behavior that is learned by the ML phase under diverse workload and system conditions. The greedy algorithm exploits this performance knowledge to approach the optimal packing point, \(Max_{pack}\), while determining an appropriate \(A_{max}\) configuration. To train the ML models without incurring the prohibitive cost of profiling a real \Gls{LLM}-adapter serving system, we introduce the DT, which reproduces system behavior with orders-of-magnitude faster execution and substantially lower resource consumption. Building this DT required an in-depth profiling and analysis of the dominant overheads in \Gls{LLM}-adapter serving, which we also report.

We evaluate our approach using the widely adopted vLLM framework~\cite{kwon2023efficient} with LoRA adapters~\cite{hu2022lora}. To illustrate the generality of the adapter caching problem beyond a single framework, we additionally provide a brief analysis using S-LoRA~\cite{sheng2024slora} in~\ref{appendix: SLoRA}.

In summary, we make the following contributions:
\begin{itemize}
    \item We propose a \textbf{data-driven pipeline for addressing the \textit{adapter caching problem}}, which aims to minimize the number of GPUs required to serve an anticipated workload. The approach maximizes per-GPU utilization while preventing request starvation and memory errors. The pipeline integrates a Digital Twin, an ML learning phase, and a greedy allocation algorithm. Results demonstrate that the proposed solution improves resource efficiency, reducing the number of GPUs required to sustain target workloads by an average of 60\% across evaluated scenarios.
    \item To the best of our knowledge, we introduce \textbf{the first Digital Twin for \Gls{LLM}-adapter serving}. The proposed Digital Twin operates orders of magnitude faster than the real system while accurately reproducing key performance metrics, and enables efficient generation of synthetic data to support the ML phase.
    \item We provide a \textbf{comprehensive analysis of the dominant overheads in \Gls{LLM}-adapter serving}, quantifying their interactions across diverse workloads and deriving actionable guidelines for system configuration and optimization.
\end{itemize}

\textit{Paper structure}: The remainder of this work is organized as follows. Sections~\ref{sec: Background} and~\ref{sec: Related work} present the necessary background and prior work, with the former also showing the adapter caching problem in practice. Sections~\ref{sec - Overview},~\ref{sec: Digital twin},~\ref{sec: ML modeling}, and~\ref{sec: Caching greedy algorithm} detail the components of the proposed pipeline. Section~\ref{sec: Evaluation} reports the main experimental results, while Sections~\ref{sec: Discussion} and~\ref{sec: Conclusions} provide discussion and concluding remarks.

\section{Background}\label{sec: Background}

\subsection{LLM serving}
An \Gls{LLM} serving system processes each request through two sequential phases: \emph{prefill} and \emph{decode}. During prefill, all input tokens are processed in parallel, producing intermediate attention states, commonly referred to as the KV cache, which are stored in GPU memory to avoid redundant computation during generation. In the decode phase, output tokens are generated autoregressively, while newly computed KV values are appended to memory. Generation terminates upon emission of an end-of-sequence token or when the maximum output length is reached.

Because decoding is inherently sequential and therefore not compute-intensive, serving systems improve hardware utilization by processing multiple requests in parallel through batching~\cite{recasens2025mind}. As static batching is inefficient under variable output lengths, modern systems instead adopt \emph{continuous batching}~\cite{yu2022orca, kwon2023efficient, tensorrt-llm, deepspeed-mii}. It allows requests to dynamically enter and exit the batch between decoding iterations, significantly improving throughput and latency.

At runtime, the number of active requests in the batch is primarily constrained by GPU memory, as each request maintains a growing KV cache throughout decoding. Early frameworks conservatively preallocated memory for the maximum possible output length, resulting in substantial overprovisioning. In contrast, systems such as vLLM~\cite{kwon2023efficient} and S-LoRA~\cite{sheng2024slora} employ greedy KV-cache allocation strategies that reserve memory only for a limited window of upcoming tokens, improving memory efficiency.

\subsection{(vLLM) Adapter serving}
Multiple types of \Gls{LLM}-adapters have been proposed~\cite{houlsby2019parameter, liu2022few, guo2020parameter, sung2022lst}. In this work, we focus on LoRA adapters~\cite{hu2022lora}, which remain the most widely adopted approach. LoRA introduces trainable weights into specific layers of the backbone \Gls{LLM} through two low-rank matrices. Their resulting activations are added to the corresponding backbone layer outputs, allowing the adapter to learn only the residual difference between the pretrained and target tasks. The size of a LoRA adapter is defined by the dimensionality of its low-rank latent space, known as \emph{rank}.

Modern serving systems support parallel execution of multiple adapters within the same batch through kernel-level optimizations~\cite{chen2024punica}. Nevertheless, a request can be processed only if both the backbone model and adapter weights reside in GPU memory. Adapter weights therefore reduce the memory available for requests KV cache. To alleviate this limitation, adapters are dynamically swap between CPU and GPU depending on their current usage.

In vLLM, GPU memory is statically partitioned at initialization by reserving a fixed region for adapter weights. This design limits the maximum number of simultaneously loaded adapters, denoted as \(A_{max}\). Although tunable, there is no principled methodology to choose its optimal value. Furthermore, vLLM assigns a uniform maximum memory footprint per adapter, \(S_{max}\), causing all adapters to occupy identical GPU memory space regardless of their actual size. S-LoRA~\cite{sheng2024slora} mitigates these limitations by jointly managing KV-cache and adapter-weight memory within a unified cache. However, given the archival status of its codebase and its limited adoption in practice, it is not used as the primary framework in this study. An exploratory analysis is provided in~\ref{appendix: SLoRA}.

\subsection{Illustrating the adapter caching problem}

\begin{figure}
    \centering
    \includegraphics[width=\linewidth]{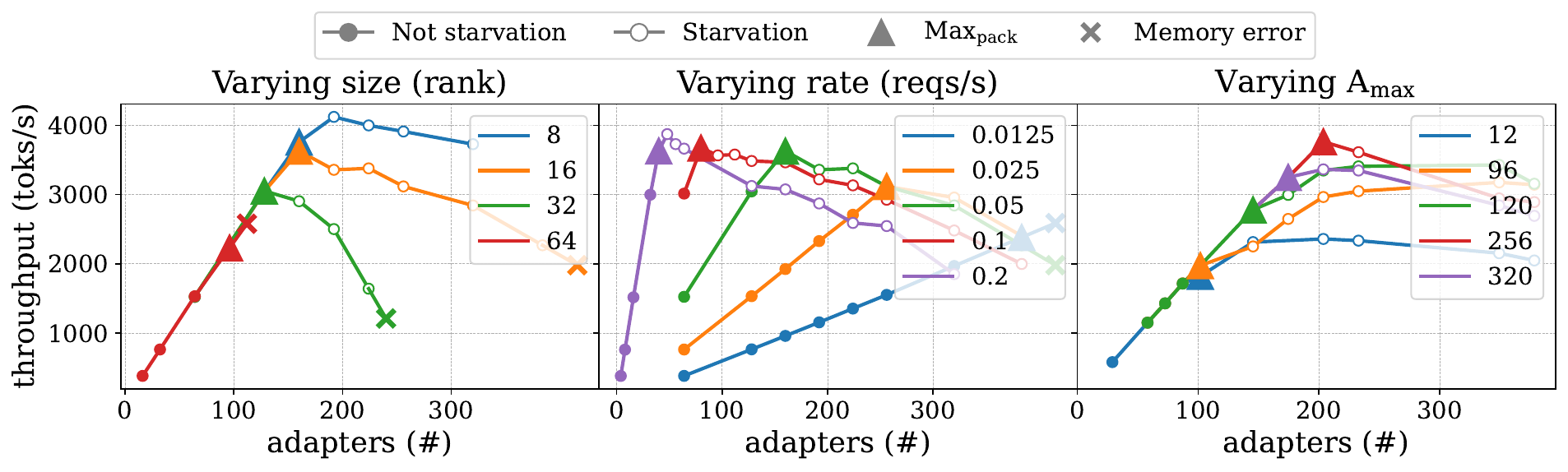}
    \caption{Throughput as a function of the number of served adapters under varying adapter sizes (left), arrival rates (center) and configured \(A_{max}\) (right). Each line corresponds to experiments in which all parameters remain fixed except for the number of adapters. Results were obtained using vLLM with Llama-2-7B~\cite{touvron2023llama2openfoundation} and a public LoRA adapter~\cite{llama27bsqlloratest} on an NVIDIA H100 Hopper GPU, processing a total of 4096 prompts submitted at the beginning of execution until completion. Default parameters were chosen to saturate the GPU between 100 and 300 adapters for improved visualization: adapter size 8, per-adapter arrival rate of 0.05 req/s, 250 input tokens per request and 231 output tokens per request. In the two leftmost plots, \(A_{max}\) is set equal to the number of served adapters and \(S_{max}\) is configured to match the adapter size used in each experiment across all plots, representing the most straightforward GPU configuration.}
    \label{fig:performance_analysis-without_offloading_variation}
\end{figure}

\begin{figure*}
    \centering
    \includegraphics[width=\linewidth]{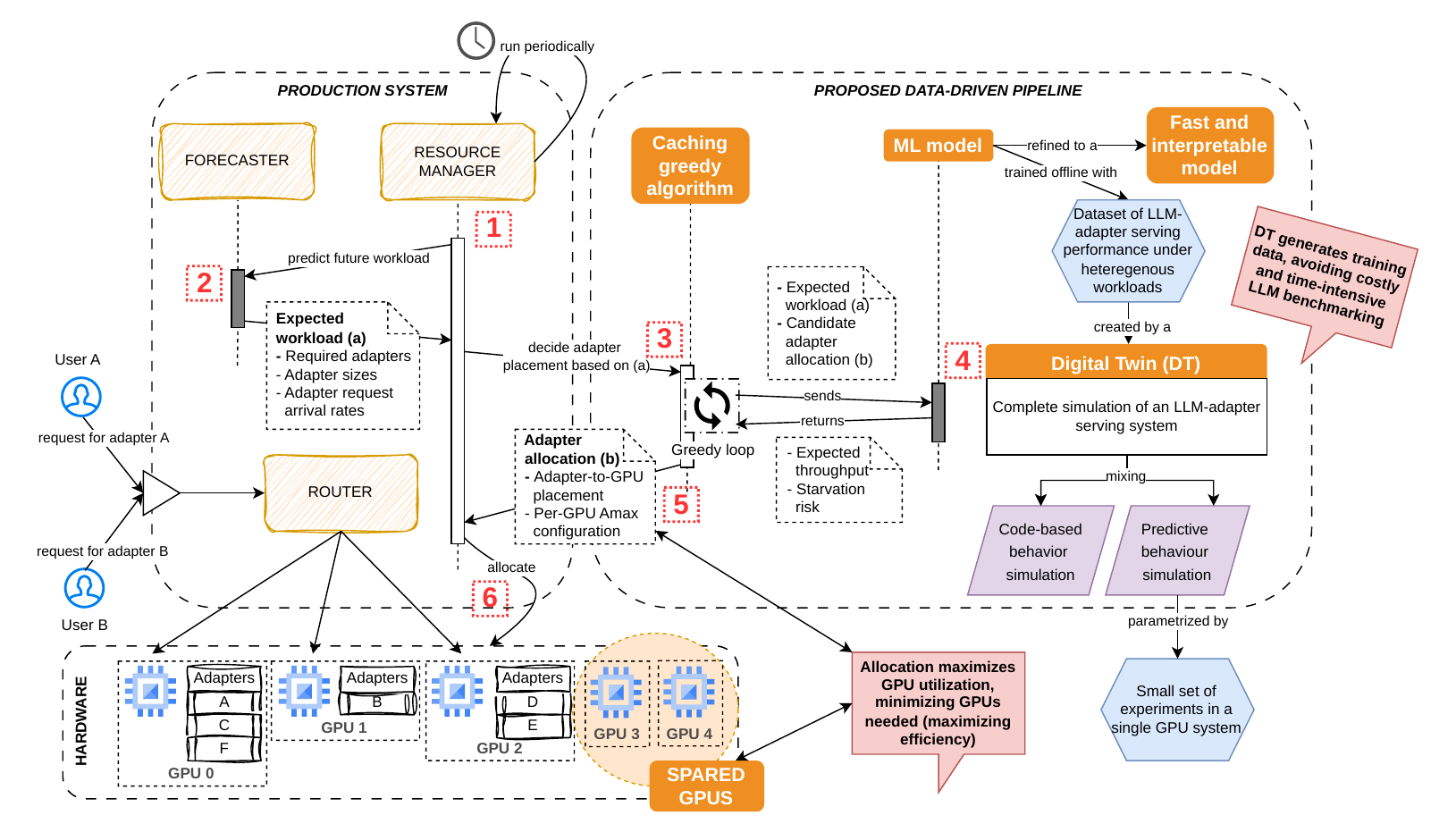}
    \caption{Proposed data-driven pipeline to address the adapter caching problem (right), shown alongside its expected usage within a production system (left). The numbered red markers indicate the recommended reading order of the workflow.}
    \label{fig:finding_maximum-diagram}
\end{figure*}

Fig.~\ref{fig:performance_analysis-without_offloading_variation} illustrates the adapter caching problem under homogeneous workloads on a single GPU. Each curve shows the throughput as a function of the number of concurrently served adapters. Two distinct regimes are consistently observed in each line. Initially, throughput increases approximately linearly as additional adapters introduce more request load that can be efficiently batched together. Beyond a certain point, throughput saturates or degrades due to insufficient GPU memory for KV-cache allocation, resulting in request starvation. The transition between these two regimes marks the optimal packing point \(Max_{pack}\) targeted in this work, which maximizes throughput while avoiding starvation. In practice, we identify this point as the highest measured throughput that remains above 90\% of the total incoming token rate.

The location of \(Max_{pack}\) is highly sensitive to adapter size, arrival rate, and the configured value of \(A_{max}\), as a consequence of the adapter-serving overheads analyzed in Section~\ref{sec: Performance Analysis}. Larger adapters reduce the available memory for request processing and increase computation cost, yielding lower \(Max_{pack}\) points. Lower arrival rates similarly obtain lower peak points, as more adapters are required to saturate GPU memory, increasing adapter overheads. The choice of \(A_{max}\) introduces a trade-off between reserving excessive memory for adapters (e.g., \(A_{max}=320\)) and restricting achievable parallelism (e.g., \(A_{max}=96\)). While the results shown correspond to homogeneous workloads, higher variance arises under heterogeneous adapter sizes and arrival rates. Lastly, memory errors (marked as crosses) occur for large adapter sizes and low arrival rates as \(A_{max}\) is trivially set equal to the number of served adapters, resulting in the reservation of more GPU memory for adapter weights than is available.

\section{Related work}\label{sec: Related work}

\subsection{LLM adapter optimization}
A growing body of work has focused on improving the efficiency of \Gls{LLM}-adapter serving systems. Existing approaches can be broadly divided into three categories. First, kernel-level optimizations, such as Punica~\cite{chen2024punica}, improve execution efficiency by enabling batching across requests associated with different adapters. Second, memory-level optimizations, exemplified by S-LoRA~\cite{sheng2024slora}, extend vLLM by dynamically partitioning GPU memory across both requests and adapters. Third, scheduling techniques, which include methods for request scheduling, such as Toppings~\cite{li2025toppings} and Chameleon~\cite{iliakopoulou2024chameleon}, as well as adapter placement strategies, discussed in the following section.

Toppings mitigates cold-start overheads and optimizes scheduling for heterogeneous adapter sizes, whereas Chameleon improves adapter cache management and introduces an express-lane mechanism for short requests. Both systems provide detailed performance analyses: Toppings primarily studies the computational overhead introduced by mixing adapters, while Chameleon focuses on \Gls{TTFT} and adapter loading latency. We extend these performance analyses by systematically studying the primary overheads of adapters and their impact on throughput and \Gls{ITL}, which form the foundation for the design of our proposed DT.

\subsubsection{Placement methods}
dLoRA~\cite{wu2024dlora} and LoRAServe~\cite{jaiswal2025serving} are prominent approaches to adapter placement. Both propose proactive placement strategies for distributed systems, analogous to the adapter caching problem, leveraging estimated long-term workload patterns derived from production traces to minimize latency and maximize throughput. dLoRA uses a heuristic based on the ratio between available GPU memory after loading adapter weights and the estimated adapter load, whereas LoRAServe uses a greedy algorithm that reduces adapter size heterogeneity by leveraging profiled single-adapter serving throughput as an estimate of peak GPU capacity. Additionally, dLoRA incorporates a reactive \Gls{ILP} based optimization to adapt placement decisions to short-term workload variations arising from unpredictable output lengths.

While dLoRA and LoRAServe aim to fully utilize available hardware resources, our work focuses on resource efficiency by achieving near-peak GPU utilization (\(Max_{pack}\)), thereby reducing the total number of GPUs required to serve a given workload. Furthermore, unlike the heuristic-based approach of dLoRA, we adopt a data-driven methodology grounded in observed serving behavior through the DT. Compared to LoRAServe, our approach captures the dynamics of serving adapters under diverse workloads and \(A_{max}\) configurations, thereby reflecting the full complexity of multi-adapter LLM serving.

\subsection{LLM simulators or digital twins} 
Several simulators have been proposed to model default \Gls{LLM} serving behavior. Vidur~\cite{agrawal2024vidur} aims to simplify server configuration selection by training random forest models on benchmarking data to predict performance metrics. LLMServingSim~\cite{cho2024llmservingsim} emulates the continuous batching loop of LLM serving systems and is primarily intended for architecture exploration. Beyond simulators, more specific analytical performance models have also been proposed. Latency is commonly approximated as a linear function of batch size, following earlier studies on generic AI models~\cite{zhang2023shepherd, 10.1145/3341301.3359658}. In the context of adapter serving, Toppings~\cite{li2025toppings} models adapter computation latency as a linear function of batch size and either the sum or the maximum adapter size within the current batch.

To the best of our knowledge, no existing simulator or digital twin explicitly models the combined dynamics of adapter caching, KV-cache allocation, and continuous batching in multi-adapter \Gls{LLM} serving. Our Digital Twin combines code-based simulation with predictive behavior modeling. The latter relies on performance models that reuse existing analytical formulations where applicable, while incorporating additional components and refinements to accurately capture adapter-serving behavior under heterogeneous workloads.

\subsection{Extension of prior workshop version}
A preliminary version of this work was presented at the NeurIPS ML for Systems Workshop (2025)~\cite{agullo2025data_driven_ml}, where throughput maximization was addressed in single-GPU settings via \(Max_{pack}\) estimation. This manuscript substantially extends that work by generalizing the problem to distributed multi-GPU environments and introducing a greedy placement algorithm for adapter allocation. Moreover, the ML component is reformulated as a distilled surrogate of the Digital Twin (DT), enabling efficient performance prediction to guide placement under heterogeneous workloads. Finally, the experimental evaluation is significantly expanded, including validation under unpredictable arrival patterns and distributed scenarios with real trace-sampled arrivals not considered in the workshop version.

\section{Pipeline overview}\label{sec - Overview}
Fig.~\ref{fig:finding_maximum-diagram} illustrates the proposed data-driven pipeline and its expected integration within a production system. Inspired by the long-term proactive strategy of dLoRA, the pipeline is designed to be periodically reinvoked based on an anticipated workload. A workload is characterized by the required adapters, their sizes, and their request arrival rates. Given this workload, the pipeline computes an adapter-to-GPU placement together with the optimal \(A_{max}\) configuration per device. The resulting allocation maximizes GPU efficiency by driving a subset of devices to their maximum feasible packing throughput (\(Max_{pack}\)), thereby minimizing the number of GPUs required to serve the workload. The remaining devices can be reassigned to other workloads to improve overall system efficiency or powered down to reduce energy consumption.

The pipeline consists of three components: (i) a Digital Twin (DT), (ii) an ML learning phase, and (iii) a greedy adapter caching algorithm. The greedy algorithm produces the final placement decision, relying on performance predictions generated by the ML models. Trained offline, these models estimate the achievable throughput of a GPU under a given adapter placement and \(A_{max}\) configuration, and predict whether starvation may arise. An optional refinement phase can simplify the models into faster and more interpretable surrogates, at the cost of limited accuracy degradation. 

Training such models requires a large and diverse dataset capturing LLM-adapter serving behavior across heterogeneous workloads. Exhaustively constructing this dataset via real-system benchmarking is computationally prohibitive due to its execution time and resource cost. To overcome this limitation, we introduce a Digital Twin that emulates the internal dynamics of an LLM-adapter serving system through a combination of code-based and predictive behavior simulation. The DT operates orders of magnitude faster and with substantially lower resource consumption than full-system benchmarking, enabling large-scale synthetic data generation to train the ML models. While the DT provides high-fidelity performance estimates, it is not directly invoked by the greedy caching algorithm, as the ML surrogate enables significantly faster inference and improved interpretability, making the pipeline suitable for production deployment. Prior to use, the DT requires a lightweight parameterization phase based on a small set of benchmarking experiments executed on the target hardware and model configuration.

\begin{figure}
    \centering
    \includegraphics[width=\linewidth, trim=0 5 0 5, clip]{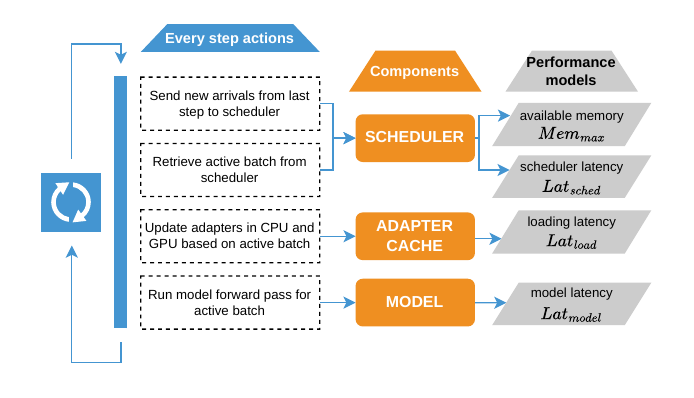}
    \caption{Digital Twin behavior and architecture.}
    \label{fig:digital_twin-architecture_diagram}
\end{figure}

\section{Digital twin}\label{sec: Digital twin}
The proposed Digital Twin (DT) reproduces an online \Gls{LLM}–adapter serving system. Rather than acting as a simple performance metric estimator, the DT implements the code loop corresponding to the continuous batching process characteristic of modern serving frameworks. This design enables accurate performance estimation under heterogeneous workload distributions. The DT operates offline and exclusively on CPU to facilitate cost-efficient generation of training data for the ML learning phase. 

Each iteration of the implemented loop is modeled as a state transition function, where the system state, comprising pending and running requests, loaded adapters, and available GPU memory, is programmatically coded and updated. These transitions are implemented by three components (scheduler, adapter cache, and model), depicted in Fig.~3, which replicate the core logic of the real system. These components integrate lightweight predictive performance models (detailed in Section~\ref{subsection: DT - Estimators}) that provide latency estimates for the time-intensive tasks of each step (i.e., model forward pass). This design enables the DT to approximate the temporal evolution of the system without executing the underlying computationally intensive tasks, thereby achieving significantly faster execution than a real system and eliminating the need for GPU resources.

As depicted in Fig.~\ref{fig:digital_twin-architecture_diagram}, each simulation step begins by injecting new request arrivals according to the simulated time and the workload arrival distribution. Requests are then passed to the scheduler, which updates the active batch by removing completed requests and admitting new ones. Following the design of vLLM, a greedy KV-cache allocation strategy is applied. The updated batch is subsequently processed by the adapter cache and model components, which emulate adapter loading/unloading as well as the model forward pass, respectively.

As a standard Digital Twin, the DT requires the same inputs as the real system to perform the simulation. These include the execution duration and detailed workload characteristics, namely the arrival time of each request, the target adapter, adapter size, request input length, and the configured GPU \(A_{max}\) value. Unlike the real system, however, the DT does not internally derive the amount of output tokens per request. Instead, the expected output length must also be provided as an input parameter. Nevertheless, our evaluation shows that using the average output length across requests yields sufficiently accurate performance estimates while remaining practical to approximate in production environments. By reproducing all major execution stages of a real serving system, the DT can generate a wide range of performance metrics, including throughput, \Gls{ITL}, and \Gls{TTFT}.

\subsection{Performance analysis}\label{sec: Performance Analysis}
We describe the four main overheads that we encountered when working with adapters. This exploration represents the foundation for the design and implementation of the DT and its predictive performance models. Experiments setup is described in Section~\ref{subsection: Methodology - Setup}.

\subsubsection{Increased memory usage}\label{subsubsection: Performance analysis - Overheads - Memory}
\begin{figure}
    \centering
    \includegraphics[width=\linewidth]{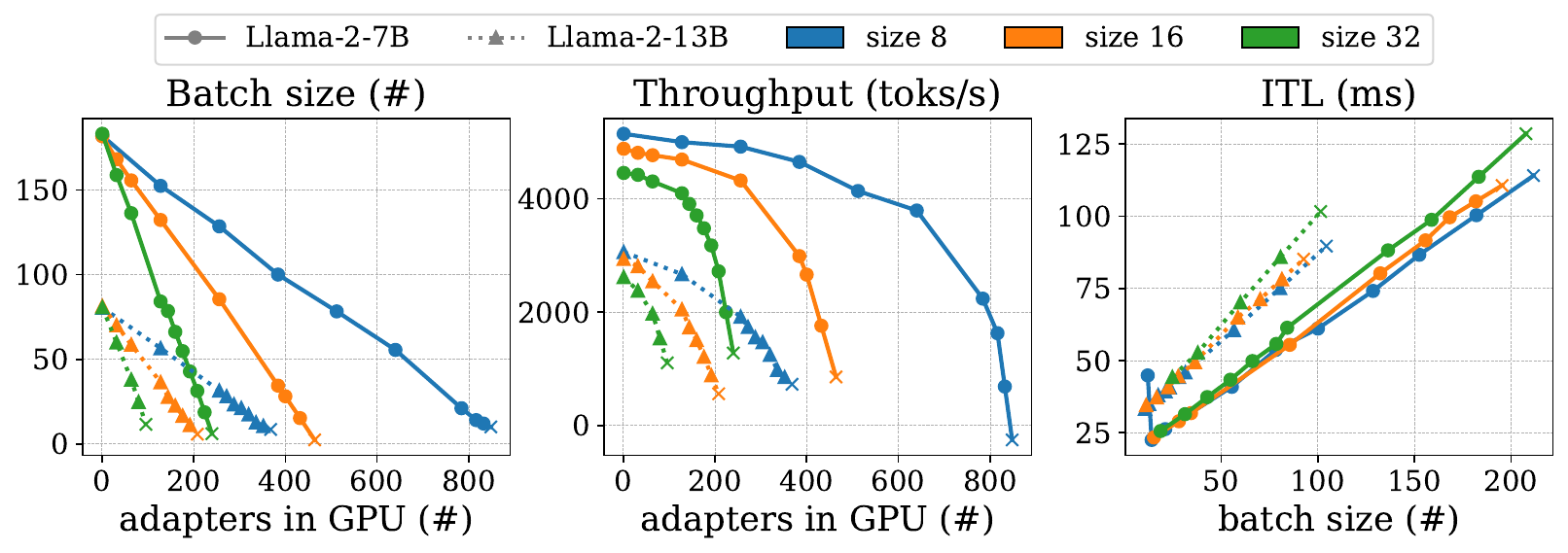}
    \caption{Evolution of batch size and throughput with increasing numbers of loaded adapters (left, center) and ITL versus batch size (right), across models and adapter sizes. Crosses denote the point where GPU memory is exhausted. Measurements were obtained by oversaturating a single-GPU system and issuing backbone-only requests to isolate the memory overhead of adapter weights. Minor variations in \gls{ITL} across adapter sizes are nevertheless observed, likely due to additional operations triggered in vLLM when adapters are activated, even if unused.}
    \label{fig:performance_analysis-memory_overhead_full}
\end{figure}

As introduced, storing adapter weights in GPU memory reduces the capacity available for the requests KV-cache, which limits the batch size and thus the throughput. Fig.~\ref{fig:performance_analysis-memory_overhead_full} shows this decrease in both batch size and throughput, which appears earlier and more sharply for larger models and adapters due to their higher memory footprint. Unexpectedly, throughput decays exponentially rather than linearly as batch size. This discrepancy arises from a broader trait of \Gls{LLM} serving unrelated to adapter processing. As reported in several works, increasing the batch size beyond a certain point leads to diminishing returns in throughput\textemdash a phenomenon known as the throughput plateau~\cite{recasens2025mind, recasens2024towards, agrawal2024taming}. This implies that, for example, with Llama-2-7B although the first 100 loaded adapters reduce the batch size, their impact on throughput is negligible, a characteristic that, to the best of our knowledge, we are the first to report. Lastly, the rightmost plot of Fig.~\ref{fig:performance_analysis-memory_overhead_full} presents the relationship between \Gls{ITL} and batch size, which, consistent with prior reports for generic AI models~\cite{zhang2023shepherd,10.1145/3341301.3359658}, exhibits a linear scaling trend.

\textbf{Insight.} Each loaded adapter affects throughput, depending on model and adapter size, but this impact may fade due to the throughput plateau.

\subsubsection{Increased computational workload}\label{subsubsection: Performance analysis - Overheads - Computation}
\begin{figure}
    \centering
    \includegraphics[width=\linewidth]{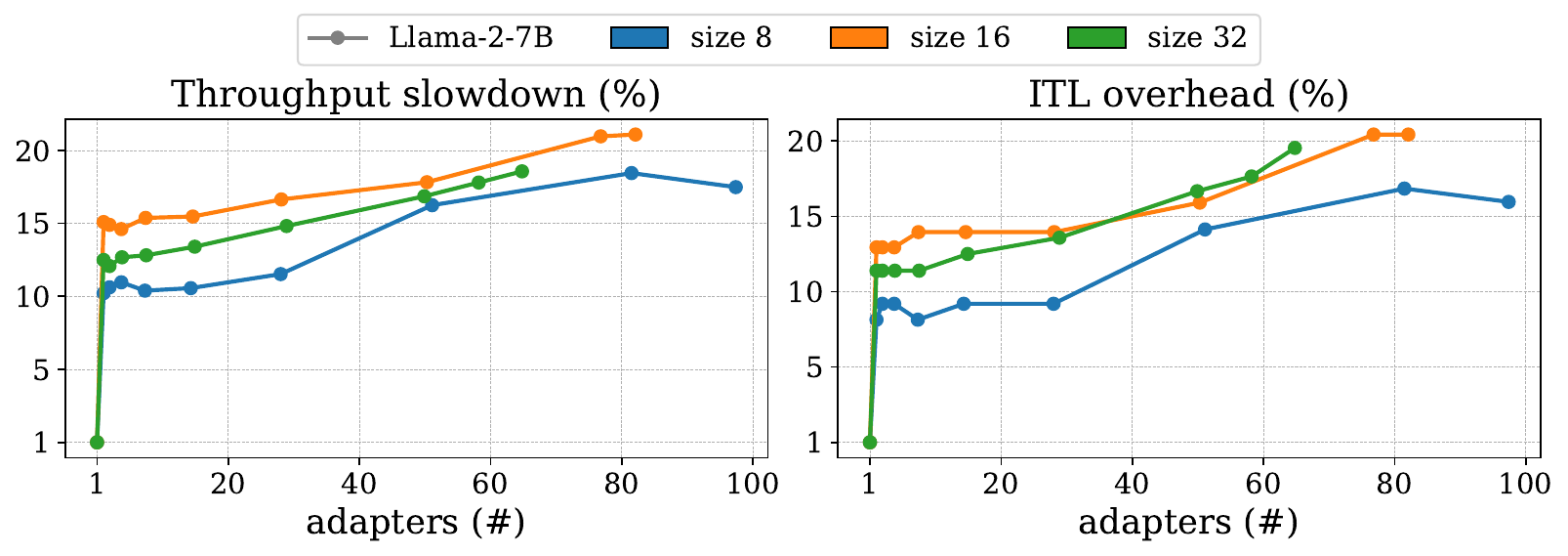}
    \caption{Throughput slowdown and \Gls{ITL} overhead for increasing adapters across three adapter sizes. Results are shown for Llama-2-7B and relative to backbone-only execution. To avoid the impact of adapter weights, we fix the batch size and number of loaded adapters within each line. Lines are shorter for larger adapter sizes due to their reduced achievable batch size, which limits the maximum number of runnable adapters.}
    \label{fig:performance_analysis-compute_overhead_full}
\end{figure}

Fig.~\ref{fig:performance_analysis-compute_overhead_full} shows the throughput slowdown and ITL overhead caused by the additional computation and GPU internal transfers required to process adapters instead of only the backbone \Gls{LLM}. Both metrics increase approximately linearly with the number of adapters, except for the sharp drop from zero to one adapter. This drop corresponds to the shift from a backbone-only execution to one that must compute both backbone and adapter activations, including the additional GPU data transfers involved. Beyond this point, computation and transfer overheads from multiple adapters is partially parallelized. Adapter size has a limited impact, with size 32 occasionally yielding lower overhead than size 16.

\textbf{Insight.} Serving adapters introduces computational overhead in both throughput and \gls{ITL} compared to backbone-only execution, increasing linearly with the number of adapters.

\subsubsection{Loading time}\label{subsubsection: Performance analysis - Overheads - Loading}
\begin{figure}
    \centering
    \includegraphics[width=\linewidth]{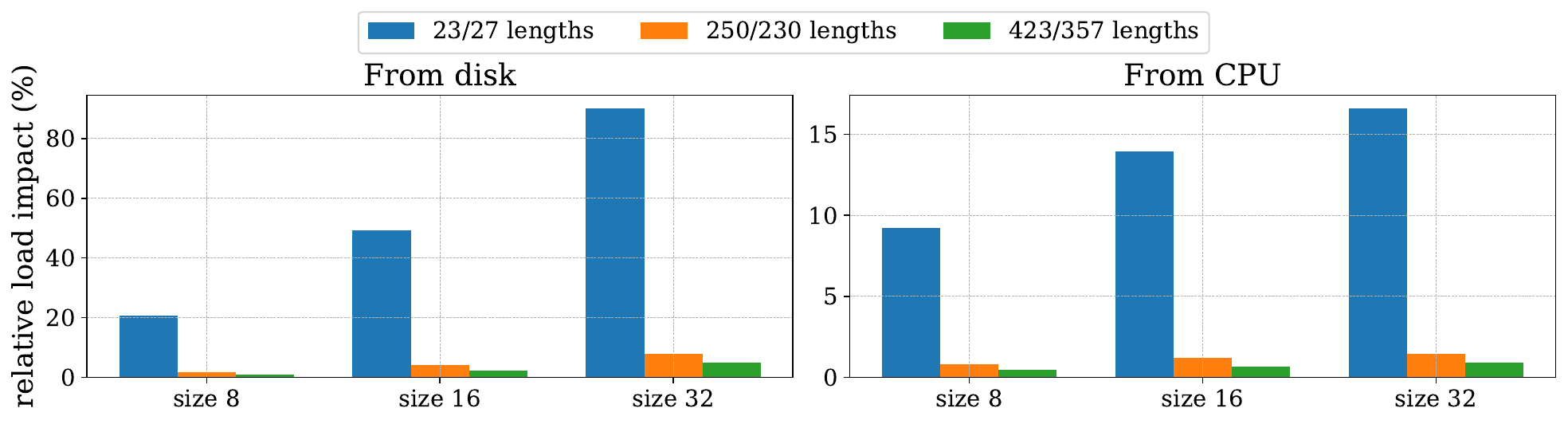}
    \caption{Loading times for varying adapter sizes, shown relative to request latency across three input/output lengths for Llama-2-7B and storage type. Request latency is computed as $TPOT * (output\_tokens - 1)$, where TPOT is the time per output token.}
    \label{fig:performance_analysis-loading_overhead_full}
\end{figure}

Fig.~\ref{fig:performance_analysis-loading_overhead_full} presents the loading time to GPU memory relative to request latency, distinguishing between loading from disk and from CPU memory. Larger adapters introduce greater overhead due to their increased size, and disk loading is on average 70\% slower than CPU loading. Request length also strongly affects the relative impact: for small requests, CPU loading adds 7–16\% latency depending on adapter size, whereas for longer requests the overhead falls below 2\%. This reduction occurs because the fixed cost of loading becomes negligible compared to the computation time of longer requests.

\textbf{Insight.} Loading overhead is significant only for short requests and can be largely mitigated by preloading adapters into CPU memory.

\subsubsection{Scheduler}\label{subsubsection: Performance analysis - Overheads - Scheduler}
\begin{figure}
    \centering
    \includegraphics[width=\linewidth]{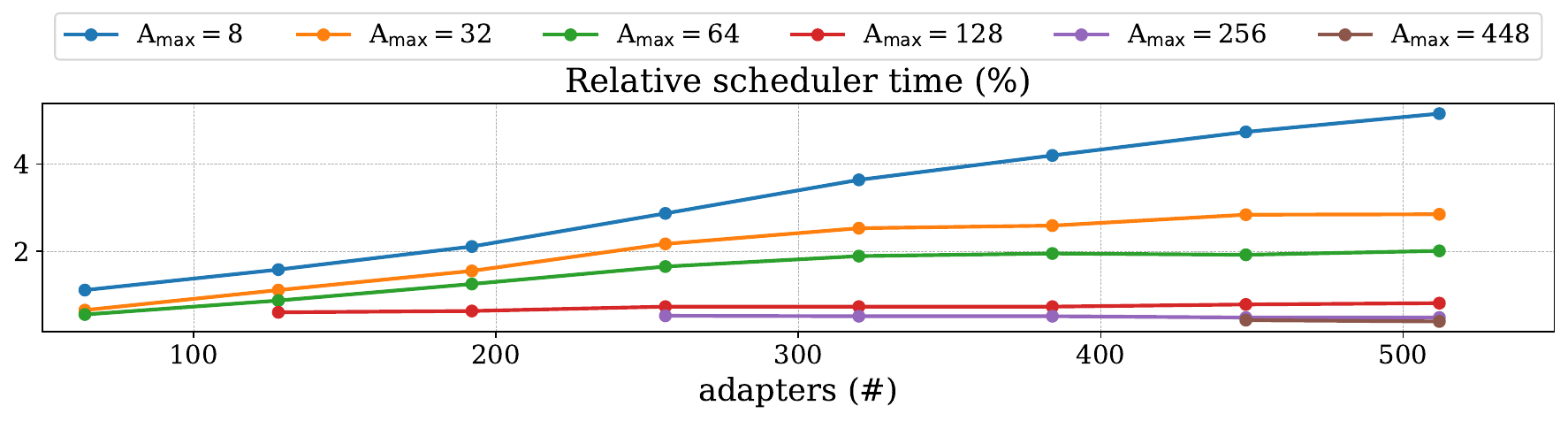}
    \caption{Scheduler time relative to the average per-step execution time, as a function of the number of adapters and configured \(A_{max}\).}
    \label{fig:performance_analysis-scheduler}
\end{figure}

Fig.~\ref{fig:performance_analysis-scheduler} shows the relative impact of the scheduler component across varying numbers of adapters and configured \(A_{max}\). When the number of adapters is high but \(A_{max}\) is small, scheduling accounts for nearly 6\% of execution time; in all other cases, its impact is negligible. This behavior is specific to the vLLM implementation rather than to \Gls{LLM} adapter serving in general. It occurs because vLLM iterates over all pending requests to select those admissible to the active batch; with small \(A_{max}\), the scheduler must scan a larger fraction of requests to find those corresponding to loaded adapters, as requests from unloaded adapters cannot be processed because \(A_{max}\) has already been reached.

\textbf{Insight.} The scheduler component of vLLM may introduce an overhead when \(A_{max}\) is small relative to the number of adapters.

\subsection{Predictive performance models}\label{subsection: DT - Estimators}
As illustrated on the right side of Fig.~\ref{fig:digital_twin-architecture_diagram}, the Digital Twin relies on four predictive performance models, summarized in Equation \eqref{eq:digital_twin-latency}, which are invoked at each step of the reproduced continuous batching loop,. Three of these models estimate task-specific latency\textemdash \(Lat_{sched}\), \(Lat_{load}\) and \(Lat_{model}\)\textemdash while the fourth, \(Mem_{max}\), serves as an auxiliary estimator used by the scheduler to determine the maximum number of requests that can be batched together under the memory pressure introduced by adapter weights (see Section~\ref{subsubsection: Performance analysis - Overheads - Memory}). Specifically, \(Mem_{max}\) takes \(A_{max}\) and \(S_{max}\) as inputs and outputs the maximum number of request tokens that can fit within GPU memory (\(T_{max}\)). This estimator is derived directly from profiled data, analogous to the leftmost plot in Fig.~\ref{fig:performance_analysis-memory_overhead_full}, but estimates the achievable token count rather than batch size. Although an analytical formulation could be derived from backbone and adapter sizes, using empirical profiling results proved more straightforward and equally accurate. 

\(Lat_{sched}\) is used by the scheduler to estimate the time spent in the original vLLM scheduling component. It takes as input the batch size (\(B\)), the number of pending requests (\(R_{P}\)), the number of unique adapters currently in the batch (\(A_{B}\)), and the total number of adapters being served (\(A\)). Its formulation follows a analytical expression parameterized using profiling data collected from the real system scheduler under diverse workload conditions. The first two terms represent the baseline scheduling behavior: an iteration over the active batch to detect completed requests and manage KV allocation, followed by a partial iteration over pending requests to assess their eligibility for inclusion. The final term models the overhead described in Section~\ref{subsubsection: Performance analysis - Overheads - Scheduler}, which accounts for the additional execution time caused by scheduler-specific inefficiencies. 

\(Lat_{load}\) estimates the time required to load adapters during swapping (\(L\)) as a function of their size \(S\) (see Section~\ref{subsubsection: Performance analysis - Overheads - Loading}), assuming unloading is negligible. Similar to \(Mem_{max}\), this model is derived directly from profiled data, specifically the measurements used to generate Fig.~\ref{fig:performance_analysis-loading_overhead_full}. Loading is modeled from CPU memory, since disk loading is treated as a one-time initialization cost.

Finally, \(Lat_{model}\) estimates the latency arising from GPU transfers and model forward pass. It is decomposed into two components: the backbone \gls{LLM} latency (\(Lat_{backbone}\)) and the additional computational overhead from serving adapters (\(Overhead_{A}\)). The backbone latency is modeled as a linear function of batch size, consistent with prior work~\cite{zhang2023shepherd, 10.1145/3341301.3359658} and supported by the profiling results in the rightmost plot of Fig.~\ref{fig:performance_analysis-memory_overhead_full}. The adapter-related overhead is modeled as a linear function of the number of adapters, following the findings in Fig.~\ref{fig:performance_analysis-compute_overhead_full}, which differs from previous modeling~\cite{li2025toppings}. The constants for both components are parametrized from the profiling data shown in the referenced figures.

\begin{equation}\label{eq:digital_twin-latency}
\begin{aligned}
    Mem_{max} (A_{max}, S_{max}) &= T_{max} \\
    Lat_{sched} (B, R_{P}, A_{B}, A) &= K_1B + K_2R_{P} + K_3R_{P}\frac{A_{B}}{A} \\
    Lat_{load} (S_{A}) &= L_A \\
    Lat_{model} (B, A) &= Lat_{backbone} * Overhead_{A} \\
                              &= (K_4B + K_5) * (K_6A + K_7)
\end{aligned}
\end{equation}

where all constants $K_x$ are parametrized with profiled data on the chosen backbone LLM, adapters, and hardware, via non-linear least squares fitting~\cite{2020SciPy-NMeth}.

\section{ML modeling}\label{sec: ML modeling}
The ML phase produces two predictive models that estimate, for a single-GPU setting, (i) achievable throughput and (ii) starvation risk for a given workload, adapter placement and \(A_{max}\) configuration. These estimators are repeatedly invoked by the greedy caching algorithm to guide distributed placement decisions. We adopt classical machine learning techniques, including Random Forests and Support Vector Machines, which provide strong predictive accuracy with moderate computational overhead. An optional refinement phase further reduces inference latency and improves interpretability, both critical for scalable and accountable production deployment.

The models are trained on a performance dataset generated by executing the DT across a wide range of workload and device configurations. Each simulated scenario contributes one sample consisting of: (i) a feature vector encoding workload and device characteristics, (ii) the DT-estimated throughput, and (iii) a binary starvation indicator. Starvation is defined as the condition where DT-estimated throughput falls below 90\% of the total incoming token rate. Although the DT supports arbitrary workload distributions, training data is restricted to long-term workload patterns that can be anticipated in advance. In practice, real production traces are approximated as per-adapter Poisson processes, with requests characterized by global average input and output token lengths, as these quantities are easier to estimate in production settings and yield sufficiently accurate results.

The feature vector characterizes both the expected workload and the GPU configuration, comprising: the number of served adapters \(A\); the sum and standard deviation of adapter Poisson arrival rates; the maximum, mean, and standard deviation of adapter sizes; and the configured \(A_{max}\) value. Two independent models are trained: a regression model for throughput prediction and a binary classifier for starvation detection.

\subsection{Refinement phase}
Among all evaluated model families, tree–based models achieve the best trade-off between predictive accuracy and interpretability. We therefore further refine these models along two complementary dimensions: inference efficiency and interpretability. Starting from the best-performing trained model, we progressively reduce model complexity until obtaining a single shallow decision tree. To this end, the hyperparameter optimization process is modified to strongly penalize complex structures. Model complexity is quantified as the number of logical decision rules of the form "\(\text{condition}_1 \land \text{condition}_2 \land \cdots \rightarrow \text{output}\)" into which a decision tree can be decomposed. After complexity reduction, inference performance is further optimized by extracting the learned decision logic and re-implementing it in plain Python, accelerated using Numba~\cite{lam2015numba}. This approach removes framework overhead and leverages code compilation to generate efficient machine code. Overall, this refinement phase substantially reduces inference latency and improves model interpretability, at the cost of a moderate degradation in predictive accuracy.

\section{Caching greedy algorithm}\label{sec: Caching greedy algorithm}
The greedy caching algorithm constitutes the final stage of the proposed pipeline. Its objective is to solve the adapter caching problem by minimizing the number of GPUs required to serve a given workload through appropriate adapter placement and \(A_{max}\) configuration, while avoiding starvation and memory errors. The algorithm leverages the ML models to estimate the expected throughput and starvation risk associated with candidate GPU allocations and device configurations. This problem can be viewed as a sophisticated variant of the bin packing problem~\cite{garey1979computers}, and is therefore NP-hard. To address it efficiently, we adopt a tailored variation of the First-Fit Decreasing (FFD) algorithm~\cite{JOHNSON1974272}, adapted to the specific constraints and objectives of the adapter caching problem.

The pseudocode is shown in Algorithm~\ref{alg:allocate-adapters}. The algorithm takes as input the expected future workload, defined by the set of adapters to serve (\(A\)), and for each adapter \(a\in A\), its expected size \(s[a]\) and Poisson arrival rate \(\lambda[a]\). It also receives the total number of GPUs in the system (\(G\)). The outputs are the GPU assignment of each adapter (\(g[a]\)) and the configuration of \(A_{max}\) per GPU (\(A_{max}[g]\)). If no starvation-free allocation is feasible, the algorithm raises an exception to enforce the non-starvation constraint. 

\begin{algorithm}
\caption{Caching greedy algorithm}
\label{alg:allocate-adapters}
\begin{algorithmic}[1]
\Require GPUs $G=\{g_1,\dots,g_M\}$; adapters $A=\{a_1,\dots,a_N\}$; sizes $s[a]$; rates $\lambda[a]$
\Ensure Assignment $g[a]$; configuration $A_{max}[g]$
\State $g[a]\gets \emptyset$ \textbf{for all} $a\in A$;\quad $A_{max}[g]\gets 0$ \textbf{for all} $g\in G$
\State $A \gets \textsc{PrioritySorting}(A,\text{key}=(\lambda[a], s[a]))$
\State $A_q \gets \textsc{Queue}(A)$;\quad $G_q \gets \textsc{Queue}(G)$
\While{$A_q\neq \emptyset$}
  \State $a \gets \textsc{PopLeft}(A_q)$
  \State \algorithmicif\ $G_q=\emptyset$ \algorithmicthen\ \textsc{Error} \textsc{Starvation}
  \State $g \gets \textsc{PopLeft}(G_q)$
  \State \textsc{ProvisionalInclude}$(g,a,s[a],\lambda[a])$
  \If{\textsc{ReachTestingPoint}$(g)$}
    \State $(ok,\;alloc\_set,\;p_{\text{new}}) \gets \textsc{TestAllocation}(g)$
    \If{$ok$}
      \State \textsc{CommitAllocation}$(g)$
      \State \ForAllDo{$a' \in alloc\_set$}{$g[a'] \gets g$}
      \State $A_{max}[g] \gets p_{\text{new}}$
      \State \textsc{PushLeft}$(G_q, g)$
    \Else
      \State $un\_alloc\_set \gets \textsc{RollbackAllocation}(g)$
      \State $A_q \gets \textsc{Merge}(A_q,\,un\_alloc\_set)$
    \EndIf
  \Else 
      \State \textsc{PushLeft}$(G_q, g)$
  \EndIf
\EndWhile
\ForAll{$g\in G$ \textbf{with} non-tested allocation}
  \State $(ok,\;alloc\_set,\;p_{\text{new}}) \gets \textsc{TestAllocation}(g)$
  \State \algorithmicif\ $not \ ok$ \algorithmicthen\ \textsc{Error} \textsc{Starvation}
  \State repeat lines 12-14
\EndFor
\State \Return $(\text{Assignment }g[a], \text{configuration }A_{max}[g])$
\end{algorithmic}
\end{algorithm}

The algorithm follows a greedy strategy, allocating adapters sequentially to maximize GPU packing up to the optimal point \(Max_{pack}\). Allocation order is determined by the \textsc{PrioritySorting} step: adapters are first sorted by size (largest first) and then by arrival rate in a zigzag order (alternating between high and low) while preserving size-based ordering. Sorting by size groups adapters with similar sizes and prioritizes larger ones first, thereby preventing newly allocated adapters from increasing \(S_{max}\) device configuration. The zigzag ordering was selected empirically, as it consistently improved throughput in our experiments. Exhaustive per-adapter testing is infeasible, so the algorithm performs provisional allocations via \textsc{ProvisionalInclude} until a predefined testing point is reached (\textsc{ReachTestingPoint}). At that point, feasibility is evaluated: successful allocations are committed (\textsc{CommitAllocation}), while failed ones are rolled back (\textsc{RollbackAllocation}), merged back into the pending queue (\textsc{Merge}), and retried on another GPU. In this work, testing points correspond to cumulative adapter counts [8, 16, 32, 64, 96, 128, 160, 192, 256, 320, 384]. After the main loop terminates, any remaining provisional allocations are validated and committed.

Feasibility is evaluated by the \textsc{TestAllocation} method, which also determines the optimal \(A_{max}\) configuration. Its pseudocode is shown in Algorithm~\ref{alg:test-allocation}. The method receives as input the GPU under evaluation and its internal state, including the number of firmly allocated adapters, the set of provisionally included adapters, and their respective sizes and arrival rates. For brevity, internal state management is omitted from the pseudocodes. The method first queries the ML model for throughput predictions using \textsc{MLPredictThroughput}, evaluating the GPU state under two configurations: the current \(A_{max}\) and the next candidate. As with adapter allocation, exhaustive exploration of all configurations is intractable. Instead, candidate \(A_{max}\) values are restricted to a predefined set returned by \textsc{NextGPUConfig}, which reuses the same array of adapter counts defined earlier. The configuration yielding the highest predicted throughput is selected. Subsequently, the method invokes \textsc{MLPredictStarvation} to check whether the selected configuration, given the current GPU internal state, would lead to starvation. If starvation is predicted, the allocation is rejected as infeasible; otherwise, the method returns the set of new allocatable adapters along with the chosen \(A_{max}\) configuration.

\begin{algorithm}
\caption{\textsc{TestAllocation}$(g)$}
\label{alg:test-allocation}
\begin{algorithmic}[1]
\Require GPU $g$; internal state $=(\mathcal{A}_{\text{alloc}},\mathcal{A}_{\text{prov}}, s[\cdot], \lambda[\cdot], p)$
\Ensure $(ok,\;alloc\_set,\;p_{\text{new}})$
\State $p_{\text{next}} \gets \textsc{NextGPUConfig}(g)$
\State $\mathcal{A}_{\text{all}} \gets \mathcal{A}_{\text{alloc}} \cup \mathcal{A}_{\text{prov}}$
\State $T \gets \textsc{MLPredictThroughput}(\mathcal{A}_{\text{all}}, s[\cdot], \lambda[\cdot], p)$
\State $T_{\text{next}} \gets \textsc{MLPredictThroughput}(\mathcal{A}_{\text{all}}, s[\cdot], \lambda[\cdot], p_{\text{next}})$
\State \algorithmicif\ $T > T_{\text{next}}$ \algorithmicthen\ $p_{\text{best}} \gets p$ \algorithmicelse\ $p_{\text{best}} \gets p_{\text{next}}$
\State $starve \gets \textsc{MLPredictStarvation}(\mathcal{A}_{\text{all}}, s[\cdot], \lambda[\cdot], p_{\text{best}})$
\State \algorithmicif\ $starve$ \algorithmicthen\ \Return $(\textsc{False}, \emptyset, \emptyset)$
\State \Return $(\textsc{True},\,\mathcal{A}_{\text{prov}},\,p_{\text{best}})$
\end{algorithmic}
\end{algorithm}

\section{Evaluation}\label{sec: Evaluation}
This section evaluates the Digital Twin and ML learning phase in approximating real \Gls{LLM}-adapter serving behavior, and the effectiveness of the caching greedy algorithm in addressing the adapter caching problem.

\subsection{Setup}\label{subsection: Methodology - Setup}
\textbf{\textit{Data.}} We sample requests from a cleaned version of the ShareGPT dataset~\cite{ShareGPT_Vicuna_unfiltered}, preserving their original heterogeneous input and output lengths. Nevertheless, for the data used to parametrize the DT performance models, we generate synthetic requests composed of random words to avoid bias in request content.

\textbf{\textit{Metrics.}} In all textual references, figures, and tables, \textit{throughput/through.} denotes the total processing rate, computed as the sum of input throughput (tokens processed as input) and output throughput (tokens generated as output).

\textbf{\textit{LLM models.}} We use Llama-3.1-8B-Instruct~\cite{grattafiori2024llama3herdmodels} and Qwen2.5-7B-Instruct~\cite{qwen2025qwen25technicalreport} with LoRA adapters of varying sizes derived from two HuggingFace adapters~\cite{llama_3.1_8B_Instruct_Finance_lora_adapter, flowertune-medical-lora-qwen2.5-7b-instruct}. As of Section~\ref{sec: Performance Analysis}, developed early in this work, we test Llama-2-7B and Llama-2-13B models~\cite{touvron2023llama2openfoundation} with LoRA adapters also based on a HuggingFace adapter~\cite{llama27bsqlloratest}. In all experiments, we set \(S_{max}\) as the maximum adapter size across each scenario, following the default behavior in vLLM.

\textbf{\textit{ML learning phase.}} We evaluate three model types: K-Nearest Neighbors (KNN), Random Forest (RF), and Support Vector Machine (SVM), all coming from the Python scikit-learn library~\cite{scikitlearn}. Training is performed with 5-fold cross-validation, and hyperparameter tuning is conducted using the HalvingGridSearchCV method from scikit-learn. The details of the hyperparameter search space are reported in~\ref{appendix:ML hyperparameter search}.

\textbf{\textit{Framework.}} Experiments use vLLM version \textit{v0.8.5}, except for the analyses in Section~\ref{sec: Performance Analysis}, which was developed earlier using version \textit{v0.5.0.post1}. In terms of implementation, we introduce minor modifications to the server components to collect auxiliary metrics and support additional arguments, and substantially update the benchmarking script to execute realistic online environments with adapters. In distributed scenarios, a different vLLM instance is deployed in each GPU with the adapter allocation provided by the placement algorithms, and requests routed according to their assigned adapter.

\textbf{\textit{Hardware.}} Experiments are conducted on a node equipped with four NVIDIA Hopper H100 (64GB HBM2), 512GB RAM memory, and 80 CPU cores.

\subsection{Digital Twin}
We evaluate the accuracy of the proposed DT by comparing its predicted performance metrics against those of a real \Gls{LLM}-adapter serving system, assessing its ability to reproduce system behavior across a wide range of different workloads. These are generated through a Cartesian combination of two adapter-size sets\textemdash high-to-low \(\{8,16,32\}\) and medium-to-low \(\{8,16\}\)\textemdash and two Poisson arrival-rate regimes\textemdash high \(\{1.6, 0.8, 0.4\}\) and low \(\{0.1, 0.05, 0.025\}\) requests per second. For each workload, adapters randomly select a size and a arrival rate from the chosen sets, producing heterogeneous configurations. The number of served adapters is varied from 8 to 384, and the corresponding values of \(A_{max}\) are explored within the same range. Real-system experiments are executed for one hour per configuration, and accuracy is quantified using the \Gls{SMAPE} across all scenarios. Results are summarized in Table~\ref{table:evaluation_table_dt_smape} under the \textit{Predictable arrivals} setting. Two variants of DT input information regarding request lengths are evaluated. In the \textit{Original} variant, the DT uses exact input and output token lengths from the real system. In the \textit{Mean} variant, all requests use identical token lengths equal to the workload average. This latter configuration reflects the information typically available in practice and corresponds to the setup used during the ML learning phase.

Overall, the DT accurately reproduces real-system behavior under predictable workloads for both models. Throughput and \Gls{ITL} achieve maximum SMAPE values of 5.08\% and 9.63\%, respectively, while \Gls{TTFT} exhibits higher deviation, with a maximum SMAPE of 18.95\%. Although the \textit{Original} variant consistently yields lower error, the \textit{Mean} configuration remains comparably accurate, supporting its use in practical scenarios. 

\begin{figure}
    \centering
    \includegraphics[width=\linewidth]{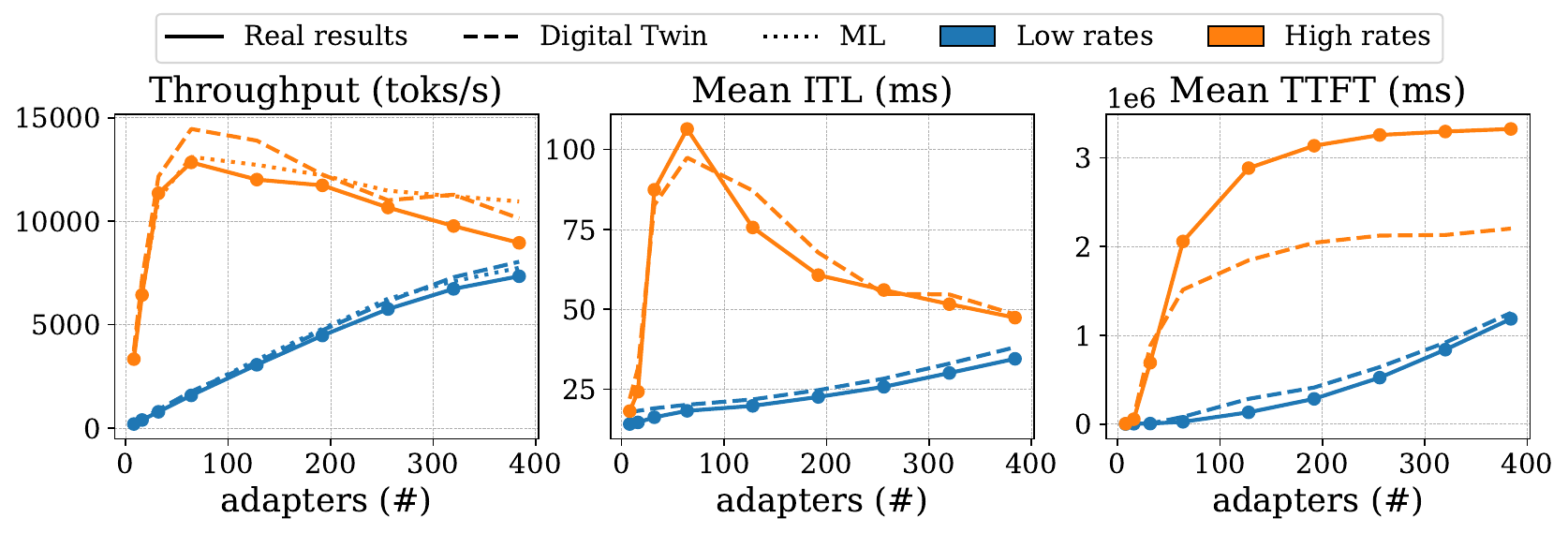}
    \caption{Comparison between DT predictions and real-system measurements for throughput, \Gls{ITL}, and \Gls{TTFT} under varying numbers of adapters and arrival rates. Experiments are conducted using adapter sizes 8 and 16 on the Qwen-2.5-7B model with \(A_{max} = 8\), in the \textit{Original} and \textit{Mean} variants. For throughput, predictions produced by the ML model are also reported.}
    \label{fig:evaluation-dt}
\end{figure}

Table~\ref{table:evaluation_table_dt_resource} shows DT time and resource usage. It achieves up to a \(90\times\) speedup compared to one-hour real-system executions, using only a single CPU core, no GPU, and approximately 200~MB of memory. Fig.~\ref{fig:evaluation-dt} illustrates a comparison between DT predictions and real-system measurements for a subset of the tested scenarios. Consistent with the table results, throughput and ITL are closely matched, whereas TTFT shows larger deviations, particularly under higher arrival rates.

\begin{table*}[!htb]
    \renewcommand{\arraystretch}{1.2}
    \centering
    \begin{tabularx}{\textwidth}{@{\extracolsep{\fill}} cc *{6}{>{\centering\arraybackslash}X}}
        \toprule
        \textbf{Model} & \textbf{Req. lengths} & \multicolumn{3}{c}{\textbf{Predictable arrivals}} & \multicolumn{3}{c}{\textbf{Unpredictable arrivals}} \\
        & & \multicolumn{3}{c}{\textbf{SMAPE comparison (\%)}} & \multicolumn{3}{c}{\textbf{SMAPE comparison (\%)}} \\        
        \cmidrule(lr){3-5}
        \cmidrule(lr){6-8}
        & & Through. & ITL & TTFT & Through. & ITL & TTFT \\
        \midrule
        \multirow{2}{*}{Llama} & Original & \textbf{2.25} & \textbf{7.09} & 18.83 & \textbf{2.47} & \textbf{5.04} & 19.95 \\
        & Mean & 4.18 & 9.63 & \textbf{17.44} & 3.66 & 9.92 & \textbf{19.91} \\
        \midrule
        \multirow{2}{*}{Qwen} & Original & \textbf{2.59} & \textbf{7.18} & 18.95 & 5.16 & 9.87 & 20.74 \\
        & Mean & 5.08 & 9.07 & \textbf{17.92} & \textbf{5.12} & \textbf{9.22} & \textbf{20.72} \\
        \bottomrule
    \end{tabularx}
    \caption{Final evaluation of the proposed DT across all predictable (left) and unpredictable (rigth) test scenarios. Reported values correspond to \Gls{SMAPE} between DT predictions and real-system measurements, where lower values denote higher fidelity.}
    \label{table:evaluation_table_dt_smape}
\end{table*}

\begin{table}[!htb]
    \renewcommand{\arraystretch}{1.2}
    \centering
    \begin{tabularx}{\columnwidth}{
        c
        *{3}{>{\centering\arraybackslash}X}
    }
        \toprule
        & \multicolumn{3}{c}{\textbf{Resource consumption}} \\
        \cmidrule(lr){2-4}
        \textbf{Model}& \mbox{Time (s)} & \mbox{CPU (\%)} & \mbox{Mem (MB)} \\
        \midrule
        Llama & $38.94\pm5.58$ & $89.51\pm1.69$ & $202.58\pm10.85$ \\
        \midrule
        Qwen & $39.61\pm5.55$ & $90.33\pm1.30$ & $204.32\pm9.29$ \\
        \bottomrule
    \end{tabularx}
    \caption{Execution time and resource consumption results of the DT across all tested scenarios.}
    \label{table:evaluation_table_dt_resource}
\end{table}

\begin{figure}
    \centering
    \includegraphics[width=\linewidth]{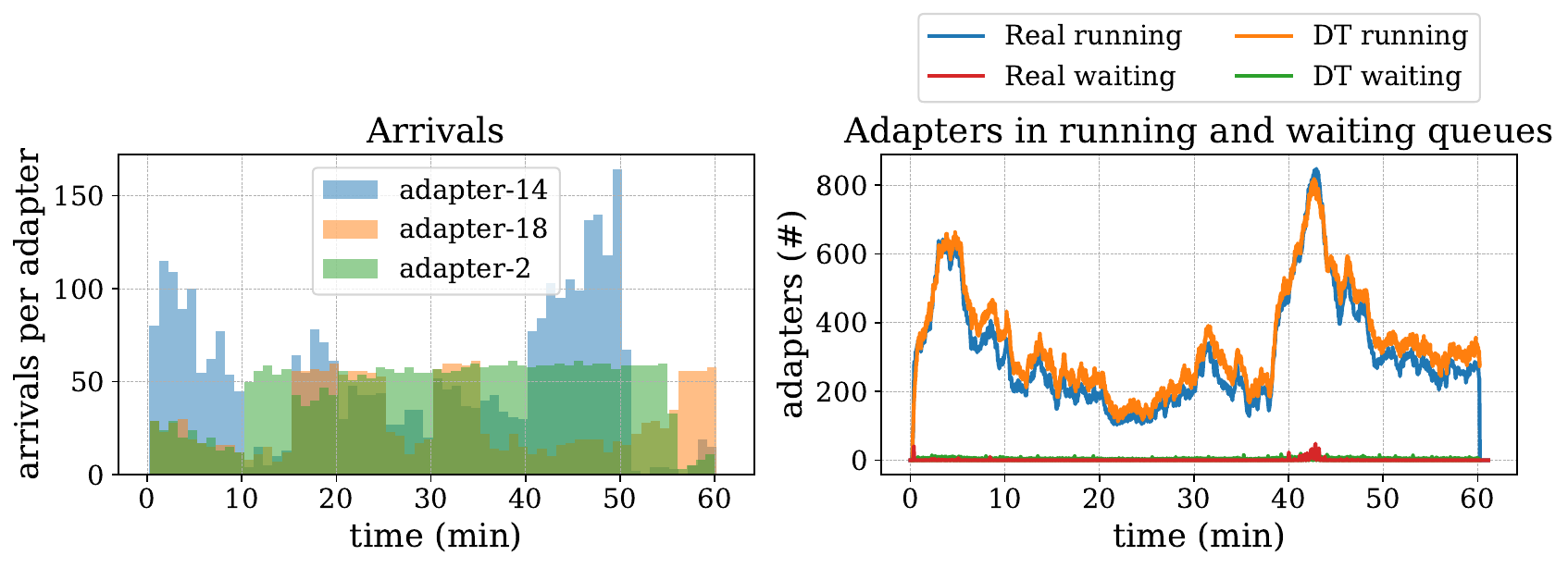}
    \caption{Execution with Llama-3.1-8B under initial high arrival rates (1.6, 0.8, 0.4) for 32 adapters in the unpredictable regime. (Left) Non-stationary arrival traces for randomly sampled adapters. (Right) Comparison of running and waiting requests over time between the DT and the real system.}
    \label{fig:dt_unpredictable}
\end{figure}

\textbf{\textit{Unpredictable arrivals}}\label{subsubsec: unpredictable}: We further evaluate the robustness of the DT under unpredictable arrivals. To that end, each adapter independently updates its arrival process every five minutes. At each update, the arrival distribution is randomly selected between Poisson and log-normal, and the corresponding arrival rate is randomly multiplied or divided by a factor of two. To prevent unrealistic behaviors, arrival rates are clipped within predefined bounds. This configuration generates strongly non-stationary traffic patterns, providing a challenging setting for assessing the DT’s ability to reproduce system behavior under rapidly changing workloads. An example of the resulting arrival traces is shown in the left plot of Fig.~\ref{fig:dt_unpredictable}.

We follow the same evaluation methodology as in the predictable-arrival experiments, where the sampled arrival rates only determine the initial values before applying the described transformations. The resulting SMAPE values are shown in the right part of Table~\ref{table:evaluation_table_dt_smape}. Overall, the DT accurately reproduces system behavior under unpredictable arrival patterns, achieving error levels comparable to those observed for predictable workloads, albeit with a slightly higher average error, which is expected given the increased complexity of the scenario. The right plot of Fig.~\ref{fig:dt_unpredictable} illustrates the evolution of running and waiting requests over time for both the DT and the real system in a specific scenario. The close alignment between the two traces highlights the DT’s ability to replicate scheduler dynamics, KV-cache allocation, adapter loading, and computation latency under highly dynamic traffic conditions.

\subsection{ML modeling}\label{subsubsec: ML model}
Leveraging the DT for rapid data generation, we train the proposed ML models over an extended and diverse set of workload scenarios. Workloads are generated through a Cartesian product of adapter sizes and arrival rates. Instead of evaluating only a small number of predefined sets, we construct all combinations of three values drawn from the adapter-size set \(\{8, 16, 32\}\) and the arrival-rate set \(\{3.2, 1.6, 0.8, 0.4, 0.1, 0.05, 0.025, 0.0125, 0.00625, 0.003125\}\). For each workload configuration, we further vary both the number of served adapters and the device configuration parameter \(A_{max}\) within the range 8 to 384. This process results in a large and heterogeneous training dataset that captures a wide range of workload conditions. The time required to generate the synthetic dataset and train the ML is reported in Table~\ref{table:evaluation_ml_resource}. Both processes exhibit execution times suitable for offline execution in production environments. Notably, the execution times for Qwen are higher, as the employed model variant is smaller than the Llama counterpart, allowing a greater number of adapters to be packed without incurring memory errors, resulting in a larger training dataset and increased processing time.

Table~\ref{table:evaluation_table_interpretable_model} summarizes the prediction accuracy of the three evaluated ML model types (KNN, RF, and SVM), evaluated against the same real-system executions employed for the validation of the DT. Throughput prediction accuracy is quantified using \Gls{SMAPE}, while starvation detection performance is evaluated using macro-averaged F1 score across all scenarios.

Overall, the ML models achieve high predictive fidelity, particularly for starvation detection. Throughput prediction errors remain below 8\%, slightly higher than those obtained using the DT, as expected given the additional abstraction introduced by the learning process. Fig.~\ref{fig:evaluation-dt} additionally reports throughput predictions produced by the best-performing RF model, showing close agreement with real-system measurements. Prediction latency, also reported in Table~\ref{table:evaluation_table_interpretable_model}, remains below 0.3~ms for all models except SVM, representing a substantial improvement over DT execution time.

\begin{table}[!htb]
    \renewcommand{\arraystretch}{1.2}
    \centering
    \begin{tabularx}{\columnwidth}{
        c
        *{3}{>{\centering\arraybackslash}X}
    }
        \toprule
        & & \multicolumn{2}{c}{\textbf{ML training time}} \\
        \cmidrule(lr){3-4}
        \textbf{Model} & \mbox{DT dataset} & \mbox{Through.} & \mbox{Starvation} \\
        & \mbox{generation} & & \\
        \midrule
        Llama & 6 hours & 270 seconds & 197 seconds \\
        \midrule
        Qwen & 9 hours & 300 seconds & 293 seconds \\
        \bottomrule
    \end{tabularx}
    \caption{Time required to generate the dataset used for training the ML models with the DT when parallelized across 80 CPUs on a single node, and time required to train the RF models for throughput and starvation prediction (including hyperparameter search), which are the ones mostly selected for online placement decisions.}
    \label{table:evaluation_ml_resource}
\end{table}

\begin{table}[!htb]
    \renewcommand{\arraystretch}{1.2}
    \centering
    \begin{tabularx}{\columnwidth}{l l *{4}{>{\centering\arraybackslash}X}}
        \toprule
        \multirow{2}{*}{\textbf{Model}} & \multirow{2}{*}{\textbf{Estimator}} 
        & \multicolumn{2}{c}{\textbf{Throughput}} 
        & \multicolumn{2}{c}{\textbf{Starvation}} \\
        \cmidrule(lr){3-4} \cmidrule(lr){5-6}
        & & SMAPE (\%) & Time (ms) & F1 (macro) & Time (ms) \\
        \midrule
        \multirow{3}{*}{Llama}
            & KNN & 4.52 & \textbf{0.21} & 0.95 & 0.19 \\
            & RF  & \textbf{4.39} & 0.25 & 0.95 & 0.16 \\
            & SVM  & 6.84 & 1.70 & \textbf{0.98} & \textbf{0.02} \\
        \midrule
        \multirow{3}{*}{Qwen}
            & KNN & \textbf{5.28} & \textbf{0.15} & \textbf{0.99} & 0.19 \\
            & RF  & \textbf{5.28} & 0.21 & \textbf{0.99} & 0.22 \\
            & SVM  & 7.46 & 2.24 & 0.93 & \textbf{0.05} \\
        \bottomrule
    \end{tabularx}
    \caption{Final evaluation results of the proposed ML model across the three model types and both LLMs. It evaluates throughput estimation (with \Gls{SMAPE}) and starvation detection (with macro-F1), and reports the average prediction time in milliseconds.}
    \label{table:evaluation_table_interpretable_model}
\end{table}

\textbf{\textit{Refinement phase}}\label{subsubsec: Explainability}: We further extend the ML learning phase with the refinement procedure described in Section~\ref{sec: ML modeling}. Table~\ref{table:evaluation_table_explainability} reports results for the original RF model, the simplified shallow decision tree (\textit{Small Tree}), and its Numba-optimized implementation (\textit{Small Tree**}).

The refinement process yields substantial improvements in both inference efficiency and interpretability. The simplified models use at most 32 decision rules, compared to at least  3.79e3 in the baseline RF. Inference latency is reduced by over \(69\times\) for the shallow tree and up to \(2120\times\) for the Numba-optimized implementation, achieving inference times below 100\,ns per prediction in some occasions. These gains are obtained at the cost of reduced predictive accuracy. On average, throughput estimation exhibits a 6.74\% increase in SMAPE, while starvation detection experiences a decrease of 0.025 in macro-F1 score. This trade-off remains acceptable in scenarios where inference speed and model interpretability are primary requirements.

Beyond performance improvements, the simplified models enable direct interpretability and extraction of actionable insights.~\ref{appendix: Simplified trees} presents the learned decision trees for both starvation and throughput prediction. For instance, we can extract that starvation is unlikely when the aggregate incoming rate remains below 23.52~tokens/s, and that excessively large \(A_{max}\) values, above 144, can negatively impact overall system throughput.

\begin{table*}[!htb]
    \renewcommand{\arraystretch}{1.2}
    \centering
    \begin{tabularx}{\textwidth}{@{\extracolsep{\fill}} l *{7}{>{\centering\arraybackslash}X}}
        \toprule
        \multirow{2}{*}{\textbf{Model}} & \multirow{2}{*}{\textbf{}} & \multirow{2}{*}{\textbf{No.}} 
        & \multicolumn{2}{c}{\textbf{Throughput}} & \multirow{2}{*}{\textbf{No.}} 
        & \multicolumn{2}{c}{\textbf{Starvation}} \\
        \cmidrule(lr){4-5} \cmidrule(lr){7-8}
        & & \textbf{rules} & SMAPE (\%) & Time (ms) & \textbf{rules} & F1 (macro) & Time (ms) \\
        \midrule
        \multirow{3}{*}{Llama}
            & RF & 7.13e6 & \textbf{4.39} & 0.25 & 4.44e4 & \textbf{0.95} & 0.16 \\
            & Small Tree & \textbf{32} & 10.25 & 3.24e-3 & \textbf{16} & 0.92 & 3.25e-3 \\
            & Small Tree** & \textbf{32} & 10.25 & \textbf{1.13e-4} & \textbf{16} & 0.92 & \textbf{8.30e-5} \\
        \midrule
        \multirow{3}{*}{Qwen}
            & RF & 6.33e6 & \textbf{5.28} & 0.21 & 3.79e3 & \textbf{0.99} & 0.22 \\
            & Small Tree & \textbf{16} & 12.89 & 2.52e-3 & \textbf{15} & 0.97 & 3.21e-3 \\
            & Small Tree** & \textbf{16} & 12.89 & \textbf{9.36e-5} & \textbf{15} & 0.97 & \textbf{10.5e-4} \\
        \bottomrule
    \end{tabularx}
    \caption{Evaluation results of the ML models obtained by progressively simplifying the best-performing RF model into a shallow, interpretable decision tree (Small Tree), and its Numba-optimized implementation (Small Tree**).}
    \label{table:evaluation_table_explainability}
\end{table*}

\subsection{Caching decisions}

\begin{figure*}[htbp]
    \centering

    \begin{subfigure}[b]{0.495\textwidth}
        \centering
        \includegraphics[width=\linewidth]{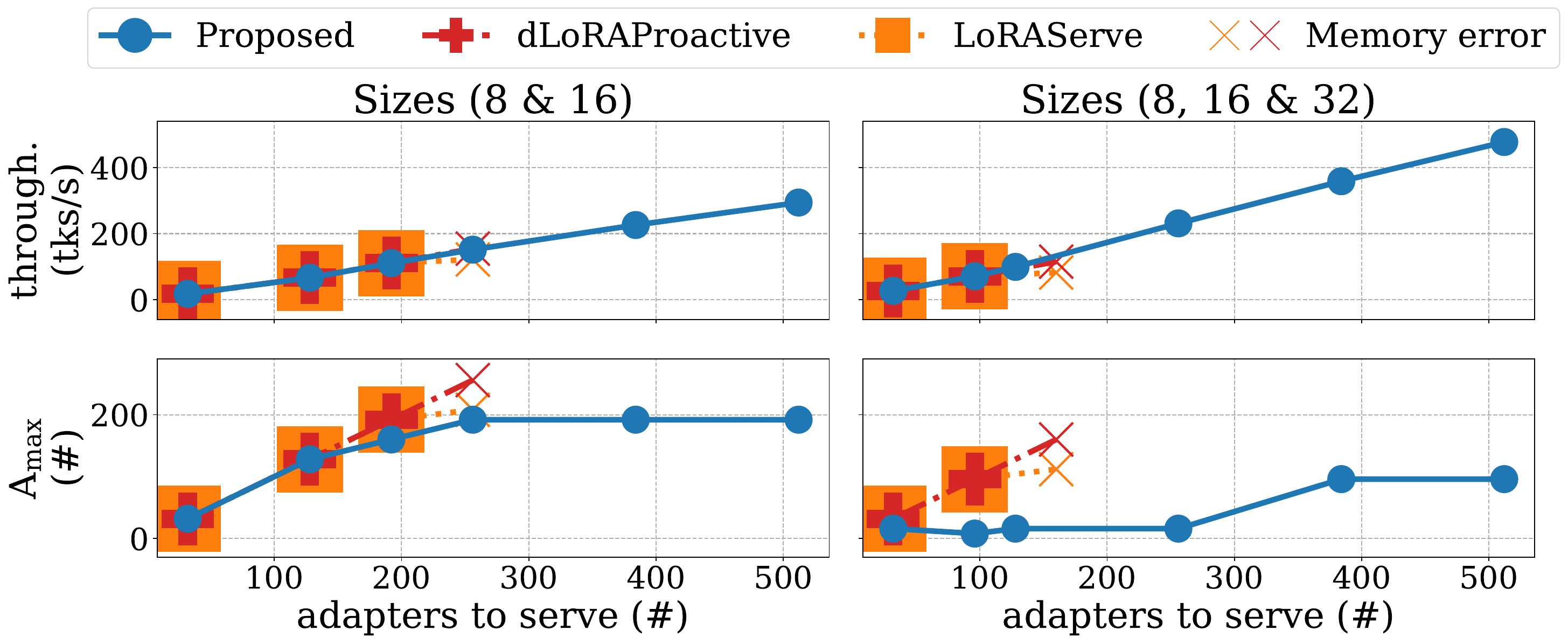}
        \caption{Llama}
        \label{fig:single_GPU_results_llama}
    \end{subfigure}
    \hfill
    \begin{subfigure}[b]{0.495\textwidth}
        \centering
        \includegraphics[width=\linewidth]{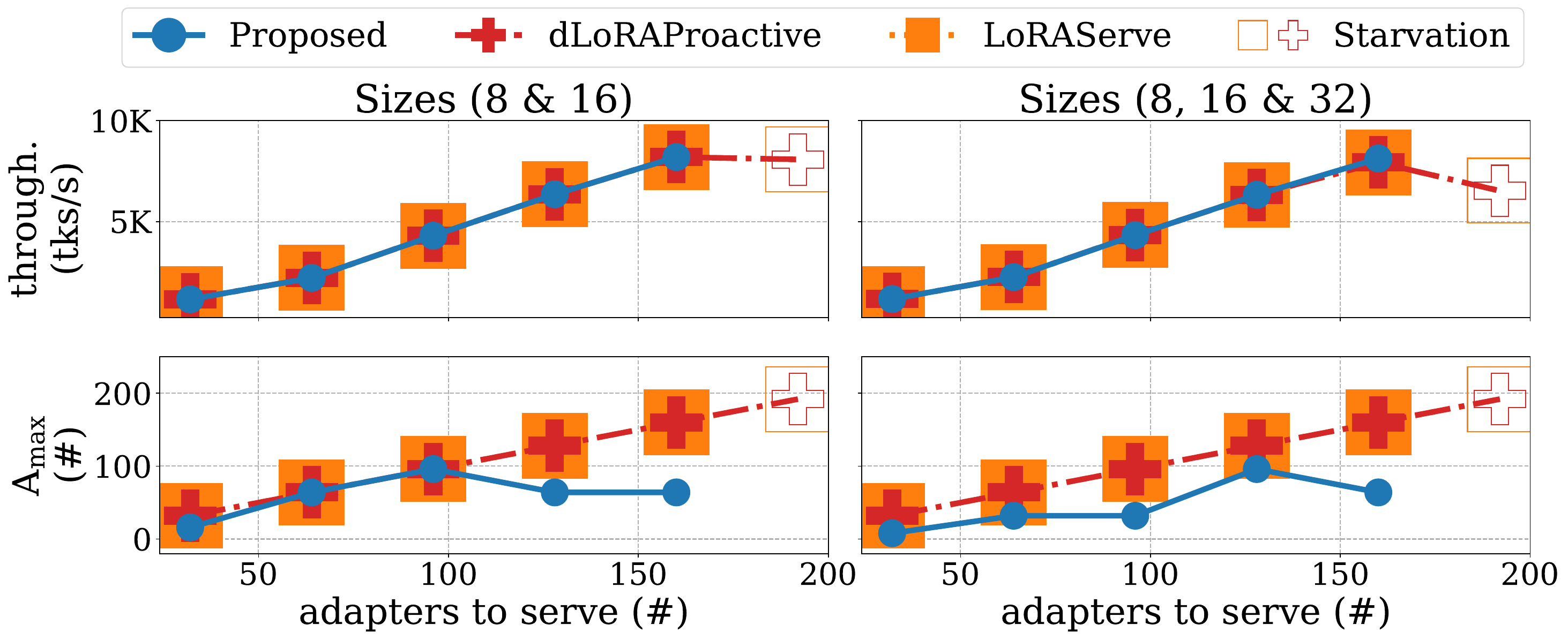}
        \caption{Qwen (arrivals $\times 100$)}
        \label{fig:single_GPU_results_qwen}
    \end{subfigure}

    \caption{(Top) Achieved throughput and (bottom) configured \(A_{max}\) on a single-GPU system by our method and the two baselines, under both adapter size settings and models. In the Qwen case, Azure arrivals are scaled by a factor of $\times 100$ to test a higher-load scenario. Each curve is shown up to the point where the corresponding method deems the placement infeasible, or encounters starvation or memory errors.}
    \label{fig:single_GPU_results}
\end{figure*}

We evaluate the proposed pipeline as a solution to the adapter caching problem, with the greedy algorithm acting as the final decision stage. We first analyze single-GPU scenarios to demonstrate throughput maximization and identification of the per-GPU optimal packing point (\(Max_{pack}\)). We then extend the analysis to distributed settings with four GPUs, showing how consistently reaching \(Max_{pack}\) improves overall GPU efficiency. Finally, we evaluate an alternative configuration of the pipeline in which the optimization objective is shifted from resource efficiency to latency minimization. 

We derive the request arrival pattern from the 2025 Azure multimodal model inference trace~\cite{qlm2024patke}, which contains data collected between October 15th and 22nd, 2024. For each subplot presented in this section, we randomly sample a single day and a pair of consecutive hours from the trace, and uniformly distribute requests across 1280 adapters. The placement algorithms are provided with the average per-adapter arrival rate observed during the first hour (i.e., modeled as Poisson process), and are tasked with determining the optimal placement for the subsequent hour. Evaluation then replays the actual arrivals from the second hour on a real \Gls{LLM}-adapter serving system using the provided placements. We consider two settings for adapter sizes: (i) sizes randomly drawn from \{8, 16\}, and (ii) sizes randomly drawn from \{8, 16, 32\}. This distinction is important, as in the latter case \(S_{max}\) must be set higher, significantly reducing the available memory for serving incoming requests.

\textbf{Baselines.} We compare our approach against the two most closely related methods:
\begin{itemize}
    \item \textbf{dLoRAProactive}~\cite{wu2024dlora}: We use the original dLoRA codebase to replicate its proactive adapter placement strategy. Other components of dLoRA are excluded, as they are orthogonal to the adapter placement problem considered in this work. When an adapter is replicated across multiple GPUs, incoming requests are randomly distributed among them. As in the original implementation, \(A_{max}\) is set to the number of adapters served per GPU, and no explicit constraint is imposed on the maximum workload that can be assigned to each GPU.
    \item \textbf{LoRAServe}~\cite{jaiswal2025serving}: As the official implementation is not publicly available, we re-implement the method based on the algorithmic description provided in the paper. When adapters are replicated across multiple GPUs, we use the output probabilities of their algorithm to distribute incoming requests accordingly. Since \(A_{max}\) is not specified, we set it to the number of adapters served per GPU, which is the most straightforward choice. Consistent with their description, we use the maximum profiled throughput for a single adapter as the upper bound on the workload assignable to each GPU.
\end{itemize}

Our approach, denoted as \textit{Proposed}, combines the caching greedy algorithm with the best-performing ML models identified in Table~\ref{table:evaluation_table_interpretable_model}, which are trained on DT data. Models obtained after the refinement phase are also evaluated in the distributed setting, labeled as \textit{ProposedFast}.

\subsubsection{\textbf{Maximizing per-GPU utilization}}\label{sec:Single-GPU system}
Figure~\ref{fig:single_GPU_results} reports the throughput achieved and configured \(A_{max}\) by the proposed placement of our method and the baselines as the number of adapters to serve increases. The left subplot show Llama under the two adapter size settings, while the right subplot presents Qwen under the same setup, with Azure trace arrivals scaled by $\times 100$ to represent a higher-load scenario.

We observe two distinct regimes. First, under low arrival rates (e.g., Llama in Figure~\ref{fig:single_GPU_results_llama}), GPU memory for adapter weights becomes the primary bottleneck as the number of adapters increases. Both baselines fail to properly control \(A_{max}\), producing placements that exceed the available GPU memory for adapter weights and result in memory errors. In contrast, our method initially allows high parallelism, with \(A_{max}\) values close to the number of served adapters, but progressively constrains it as the number of adapters grows, particularly in the setting that includes larger adapter sizes (i.e., 32). This prevents memory violations and enables serving over 500 adapters, compared to fewer than 200 for the baselines, achieving more than \(2\times\) higher throughput.

The second regime arises under high arrival rates (e.g., Qwen in Figure~\ref{fig:single_GPU_results_qwen}). In this setting, the main bottleneck shifts to the memory required for request KV values, driven by the large number of concurrent requests. As the number of adapters increases, insufficient KV capacity leads to request starvation in both baselines. Our method, however, detects such infeasible configurations and stops before generating placements that would incur starvation. Instead, it limits the per-GPU workload to the maximum feasible throughput (\(Max_{pack}\)), ensuring each GPU operates within its capacity. Excess demand is deferred to other GPUs, as discussed in the next section, thereby preventing starvation while maximizing per-GPU utilization.

\begin{figure*}[t]
    \centering
    
    \begin{subfigure}{1\linewidth}
        \centering
        \includegraphics[width=\linewidth]{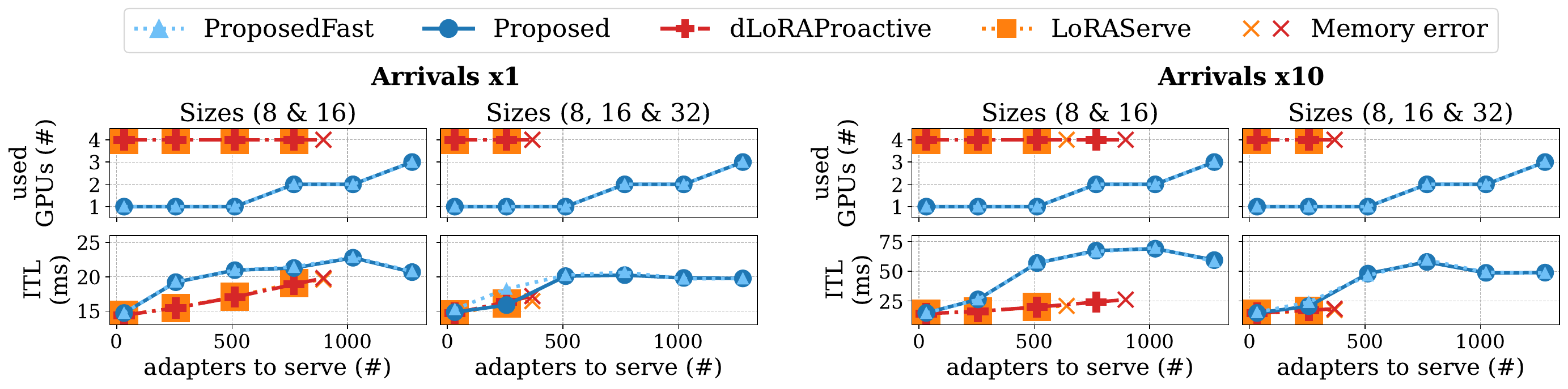}
        \caption{Llama (arrivals $\times 1$ and $\times 10$)}
        \label{fig:multi_GPU_results_llama}
    \end{subfigure}
    
    \vspace{0.5cm} 
    
    \begin{subfigure}{1\linewidth}
        \centering
        \includegraphics[width=\linewidth]{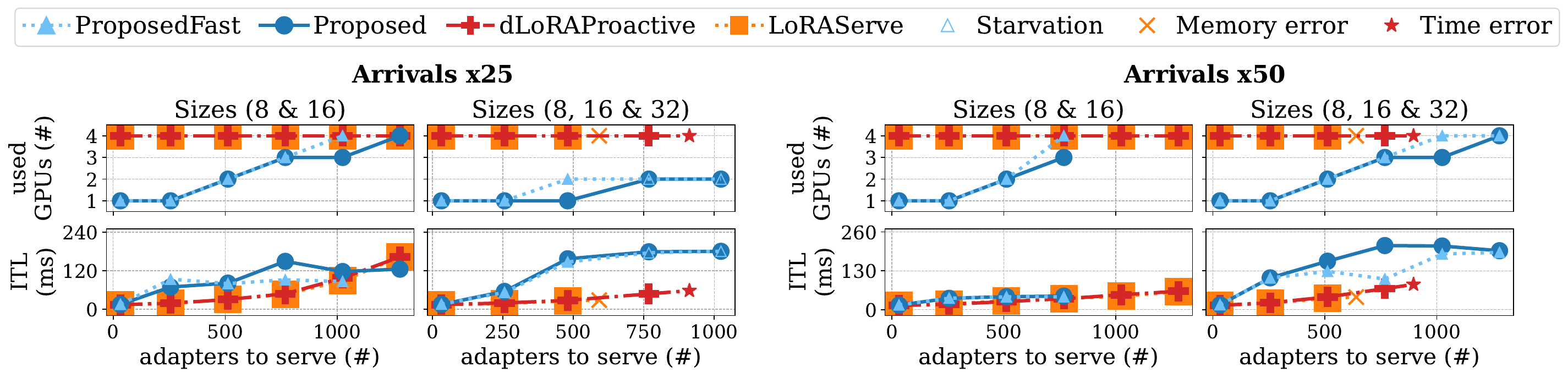}
        \caption{Qwen (arrivals $\times 25$ and $\times 50$)}
        \label{fig:multi_GPU_results_qwen}
    \end{subfigure}
    
    \caption{Number of GPUs required and \Gls{ITL} achieved by each method on a 4-GPU system, for both models under increasing arrival load as the number of served adapters scales from 32 to 1280. Each curve is shown up to the point where the corresponding method deems the placement infeasible or encounters starvation or memory errors. Time-limit failures for dLoRA occur when its placement algorithm does not complete within the imposed timeout of two hours (while the scenario evaluation itself lasts one hour). These failures arise for large adapter counts and model sizes, which may be consistent with the original work, where not evaluation was done beyond 128 adapters.}
    \label{fig:multi_GPU_results}
\end{figure*}

\subsubsection{\textbf{Efficient utilization of GPU resources}}
Figure~\ref{fig:multi_GPU_results} reports the number of GPUs utilized and the achieved ITL for our approach and the two baselines in a distributed setting with four GPUs. Results are presented for both models and adapter settings under increasing arrival load conditions (from $\times 1$ to $\times 50$), while scaling the number of served adapters from 32 to 1280, following the initial partitioning of the Azure trace.

The baseline methods exhibit memory errors, as they do not properly adapt the \(A_{max}\) configuration to the workload characteristics, consistent with the behavior previously observed in Figure~\ref{fig:single_GPU_results_llama}. As a result, they are unable to scale to higher adapter counts. In contrast, our method successfully serves a substantially larger number of adapters. For example, in the leftmost Llama subplot, it supports up to 1280 adapters, whereas the baselines are limited to 768. These limitations arise earlier for Llama (8B) than for Qwen (7B) due to the larger size of the former, which results in a higher memory footprint per adapter.

The key observation is that our method leverages estimates of near-peak GPU utilization, previously seen in Figure~\ref{fig:single_GPU_results_qwen}, to maximize adapter packing per GPU. This enables a significant reduction in the number of GPUs required to serve a given workload. While the baselines utilize all available GPUs, our approach reduces GPU usage by an average of 2.4 GPUs (60\%) across the evaluated scenarios. For instance, in the rightmost Qwen configuration, our method serves 256 and 768 adapters using only 2 and 3 GPUs (out of 4), respectively, while still serving the workload without starvation. An exception is observed in the center-right Qwen configuration, where peak utilization is underestimated, leading to incorrectly determining that workloads beyond 768 adapters cannot be accommodated within the available resources.

These efficiency gains come at the cost of increased latency, as reflected in the \Gls{ITL} metric. This is expected, as higher adapter counts increase request load and batch size, thereby raising latency, consistent with Figure~\ref{fig:performance_analysis-memory_overhead_full}. Ultimately, this trade-off depends on the desired optimization objective, in our case, the focus is on maximizing GPU efficiency rather than minimizing latency (see the following section for a latency-oriented variant of our approach).

Finally, Table~\ref{table:time_proposed_table} reports the average execution time required to generate a placement. While the \textit{Proposed} approach incurs higher computational overhead than the baselines, its latency remains suitable for periodic allocation updates. The \textit{ProposedFast} variant, which incorporates a refinement phase in the underlying ML models, achieves comparable execution cost to LoRAServe and is approximately two orders of magnitude faster than \textit{dLoRAProactive}, with an average latency of 3--4 ms per placement. Moreover, results in Figure~\ref{fig:multi_GPU_results} show that \textit{ProposedFast} maintains performance close to \textit{Proposed}, delivering a substantial improvement in GPU resource efficiency over the baselines. However, this speedup comes at the cost of less stable predictions, as illustrated by the left-center Qwen subplot, it may produce placements that incur starvation, indicating a trade-off between model performance and execution cost.

\begin{table}[!htb]
    \renewcommand{\arraystretch}{1.25}
    \setlength{\tabcolsep}{10pt}
    \centering
    \begin{tabularx}{\columnwidth}{l *{3}{>{\centering\arraybackslash}X}}
        \toprule
        \textbf{Method} & \multicolumn{2}{c}{\textbf{time(s)}} \\ 
        \cmidrule(lr){2-3}
        & Llama & Qwen \\
        \midrule
        Proposed & 1.590 & 1.785 \\
        ProposedFast & 0.004 & 0.004 \\
        LoRAServe & 0.003 & 0.004 \\
        dLoRAProactive & 0.654 & 1.013 \\
        \bottomrule
    \end{tabularx}
    \caption{Average execution time per placement for the baselines and the two variants of our method, \textit{Proposed} and \textit{ProposedFast}, in the distributed scenario.}
    \label{table:time_proposed_table}
\end{table}

\subsubsection{\textbf{Latency oriented}}\label{subsubsection: Latency oriented}
\begin{figure}
    \centering
    \includegraphics[width=\linewidth]{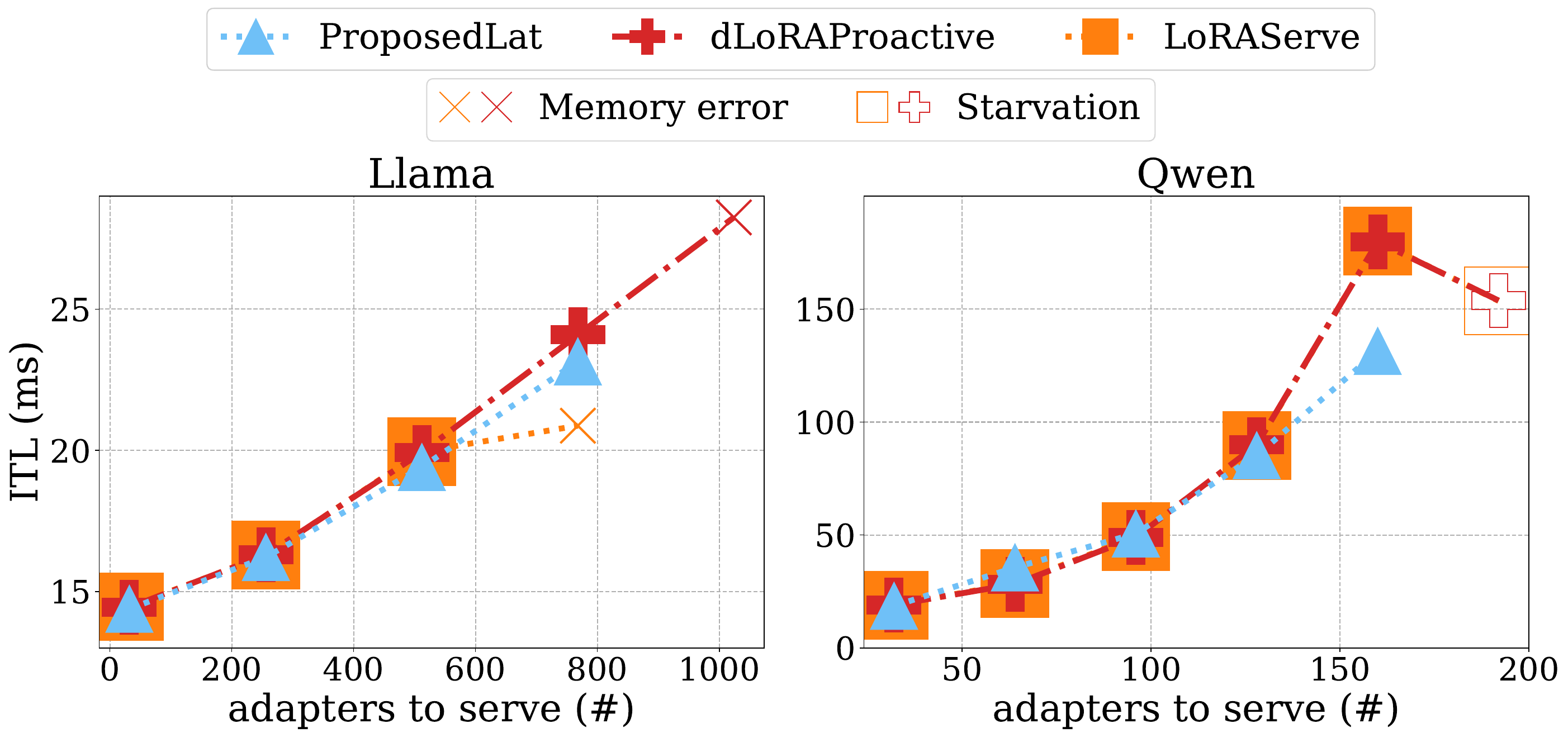}
    \caption{\Gls{ITL} obtained for two instances from the previous evaluations, comparing the latency-oriented variant \textit{ProposedLat} against the two baselines. Specifically, (left) Llama with arrivals scaled by $\times 10$ and adapter sizes 8 and 16 in the distributed scenario; and (right) Qwen with arrivals scaled by $\times 100$ and adapter sizes 8, 16, and 32 in the single-GPU scenario.}
    \label{fig:latency_results}
\end{figure}

To assess whether the proposed pipeline can be adapted to alternative objectives beyond GPU efficiency, we implement a proof-of-concept variant targeting latency minimization. This prototype, denoted \textit{ProposedLat}, reuses the learned ML models but replaces the throughput-oriented greedy algorithm with a latency-oriented heuristic. Specifically, it assigns adapters sequentially to the GPU with the lowest aggregated arrival rate and configures \(A_{max}\) as the number of adapters served on each GPU. After all adapters are assigned, the resulting allocation is validated using the learned ML models. Allocations predicted to yield a throughput lower than the incoming token rate, or to incur memory errors, are deemed infeasible. Fig.~\ref{fig:latency_results} presents its results alongside the two baselines, for two instances from the preceding evaluations.

As shown, \textit{ProposedLat} achieves latency comparable to the baselines as it also utilizes all available GPUs by design. The key distinction lies in the integration of the learned ML models, which enables to avoid infeasible allocations that lead to memory errors (left) or request starvation (right). This makes \textit{ProposedLat} more suitable for production environments and highlights the practical value of the proposed contributions beyond maximizing GPU efficiency. Notably, \textit{ProposedLat} employs a simple \(A_{max}\) configuration, which could be further improved as in the default \textit{Proposed} strategy.

\section{Discussion}\label{sec: Discussion}
Our evaluation demonstrates that the proposed pipeline significantly enhances GPU efficiency in distributed \Gls{LLM}-adapter serving systems, achieving an average reduction of 60\% in the number of required GPUs compared to baselines (Fig.~\ref{fig:multi_GPU_results}). This improvement stems from systematically identifying the \(Max_{pack}\) operating point and the corresponding \(A_{max}\) configuration that maximize per-GPU throughput while avoiding starvation and memory errors (Fig.~\ref{fig:single_GPU_results}). Consequently, the system can serve the target workloads using a reduced subset of devices, thereby freeing the remaining GPUs for additional tasks or potential energy savings. The online computation time of the pipeline, reported in Table~\ref{table:time_proposed_table}, meets the requirements of periodic reconfiguration under predictable workload conditions. Furthermore, the optimized variant, \textit{ProposedFast}, refines the underlying machine learning models into shallow, interpretable decision trees, reducing execution time by several orders of magnitude while enhancing transparency in the decision-making process. This efficiency gain is achieved at the cost of a slight increase in allocation instability, a trade-off that is acceptable in scenarios where rapid placement estimation is prioritized over predictive accuracy.

As anticipated, increasing per-GPU packing leads to higher latency compared to baseline approaches, primarily due to larger effective batch sizes. Nevertheless, we further evaluate a latency-oriented variant of the proposed pipeline, \textit{ProposedLat} that achieves latency comparable to baseline methods while maintaining robustness against starvation and memory errors. These results indicate that the proposed pipeline can also be beneficial for alternative optimization objectives beyond resource efficiency.

Aside its role in the proposed pipeline, the DT constitutes an additional contribution. To the best of our knowledge, it is the first digital twin specifically tailored to \Gls{LLM}-adapter serving. As shown in Table~\ref{table:evaluation_table_dt_smape} and Figs.~\ref{fig:evaluation-dt} and~\ref{fig:dt_unpredictable}, it closely reproduces real-system performance, particularly throughput, across both predictable and unpredictable workloads, while operating significantly faster and at substantially lower computational cost (Table~\ref{table:evaluation_table_dt_resource}). This enables not only rapid synthetic dataset generation, but also broader applications such as scheduling optimization and server configuration exploration.

Finally, Section~\ref{sec: Performance Analysis} provided an in-depth characterization of \Gls{LLM}-adapter serving behavior, detailing the impact of the four principal overheads introduced by adapters. Beyond being the basis for the design and implementation of the proposed DT, this analysis offers independent value by exposing key system-level relationships, such as the connection between the adapter memory footprint overhead and the throughput plateau.

\subsection{Limitations and future work}
Even though the evaluation is conducted using the ShareGPT dataset, which exhibits a heterogeneous distribution of input and output sequence lengths, reliance on a single dataset limits the generalization of the results to workloads with substantially different length characteristics. Future work will extend dataset generation to encompass a broader range of sequence-length distributions, enabling the placement strategy to better generalize to more diverse and heterogeneous serving conditions.

Additionally, an important direction for future research is the study of online retraining mechanisms for both the DT and the associated ML models, allowing the system to dynamically adapt to evolving workload patterns observed in production traces. Finally, we will investigate the underlying causes of the sole exception in which the \textit{Proposed} method exhibits overly conservative behavior (right-center subplot of Qwen in Fig.~\ref{fig:multi_GPU_results}), classifying certain workloads as infeasible despite being successfully served by the baselines. This limitation is likely attributable to gaps in the training data generated by the DT, which should be expanded. In particular, the initial evaluation of this work tested exclusively on Poisson arrival processes and did not account for the Azure trace characteristics considered in the current evaluation.


\section{Conclusions}\label{sec: Conclusions}
We presented a data-driven pipeline to address the adapter caching problem, improving GPU efficiency in distributed \gls{LLM}-adapter serving through workload-aware adapter placement. We evaluated the approach under heterogeneous and diverse workloads against state-of-the-art baselines. The results demonstrate that the pipeline effectively maximizes per-GPU throughput while minimizing the number of GPUs required to serve the target workloads, without incurring starvation or memory errors. The pipeline can be integrated into production systems to periodically update placements for predictable workload patterns, reducing the amount of needed hardware. The freed GPUs can be reassigned to other workloads to improve overall system efficiency or powered down to reduce energy consumption. Central to this approach is a Digital Twin for \gls{LLM}-adapter serving, which closely reproduces real-system performance at low cost, enabling efficient ML training with performance data and supporting additional optimization tasks beyond this work.

\section*{Acknowledgment}
This work has been partially financed by the EU-HORIZON MSCA programme under grant agreement EU-HORIZON MSCA GA.101086248. Also, it has been partially financed by Generalitat de Catalunya (AGAUR) under grant agreement 2021-SGR-00478, by Severo Ochoa Center of Excellence CEX-2021-001148-S-20-3, and by the Spanish Ministry of Science (MICINN), the Research State Agency (AEI) and European Regional Development Funds (ERDF/FEDER) under grant agreement PID2024-160996OB-I00, MICIU/AEI/10.13039/ 501100011033/FEDER, UE.

The authors used OpenAI’s GPT model to assist with text rephrasing and stylistic refinement. All content was originally written and subsequently reviewed and validated by the authors.

\appendix
\section{S-LoRA}\label{appendix: SLoRA}
\begin{figure}[!htb]
    \centering
    \includegraphics[width=0.8\linewidth]{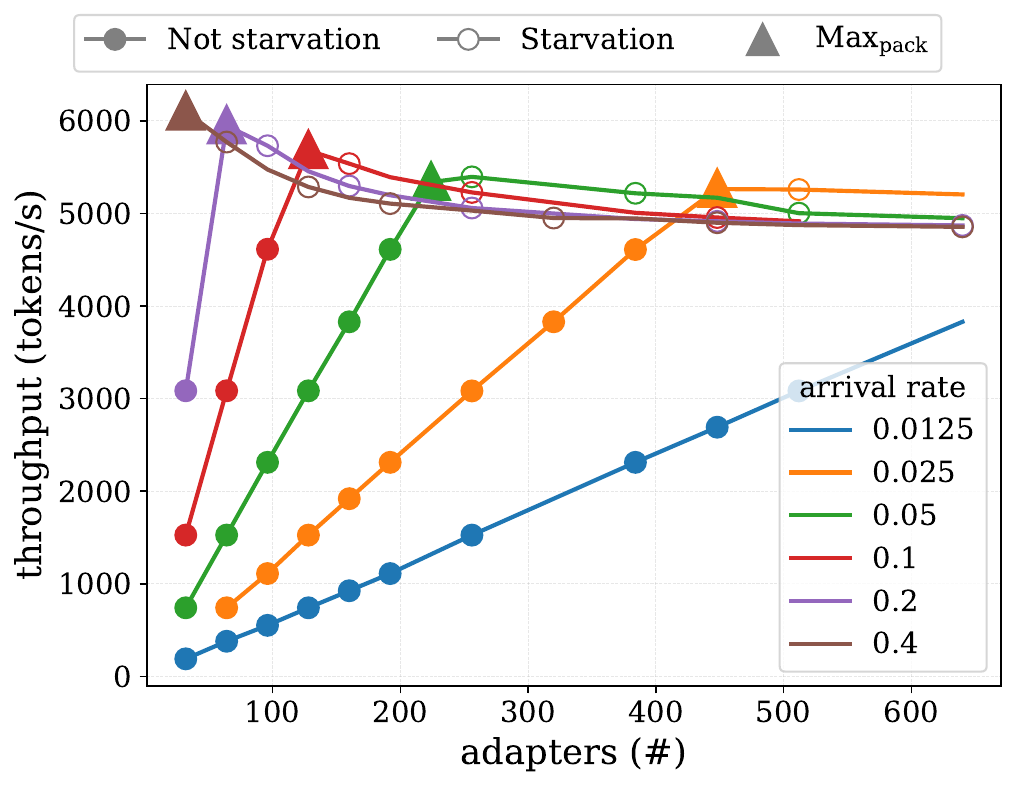}
    \caption{Throughput of S-LoRA on Llama-2-7B under varying adapter arrival rates, using 32-sized adapters and fixed request lengths (250 input tokens, 231 output tokens).}
    \label{fig:performance_analysis-slora}
\end{figure}

We include a brief analysis using the S-LoRA framework~\cite{sheng2024slora} to show that the adapter caching problem is not specific to vLLM and also arises in other serving systems. Fig.~\ref{fig:performance_analysis-slora} reports the \(Max_{pack}\) point in S-LoRA across different arrival rates, for a fixed adapter size and request-length distribution, following the same methodology as the middle plot of Fig.~\ref{fig:performance_analysis-without_offloading_variation} for vLLM. Notably, throughput degradation plateaus as the number of adapters increases, whereas vLLM exhibits a more pronounced decline, highlighting the impact of S-LoRA's design choices. Nevertheless, identifying \(Max_{pack}\) remains necessary to determine how many adapters can be allocated per GPU without triggering starvation, and the throughput level at which \(Max_{pack}\) occurs still varies across workloads, with additional variability expected when changing adapter sizes and request-length distributions.

\section{ML hyperparameter search space}\label{appendix:ML hyperparameter search}
Hyperparameter optimization in the ML learning phase is performed using \texttt{HalvingGridSearchCV} with 5-fold cross-validation. Distinct search spaces are defined for the two tasks, throughput regression and starvation classification, using the corresponding scikit-learn model classes. The evaluated values for each search are listed below.

\textbf{Throughput (Regression).} 
a) Random Forest (\texttt{RandomForestRegressor}): 
\texttt{n\_estimators} \{32,128,256\}, 
\texttt{max\_depth} \{None,5,10,20\}, 
\texttt{min\_samples\_split} \{2,5,10,20\}, 
\texttt{criterion} \{squared\_error, absolute\_error, friedman\_mse, poisson\}, 
\texttt{min\_samples\_leaf} \{1,2,5,10,32,128\}, 
\texttt{max\_features} \{auto,sqrt,log2\}.  
b) SVM (\texttt{SVR}): 
\texttt{C} \{0.1,1,10,100,1000,10000\}, 
\texttt{kernel} \{linear,poly,rbf,sigmoid\}, 
\texttt{epsilon} \{0.1,0.5,1,5\}, 
\texttt{degree} \{2,3,4,5\}, 
\texttt{gamma} \{scale,auto,0.01,0.1,1,10\}, 
\texttt{coef0} \{0,0.1,0.5,1\}.  
c) KNN (\texttt{KNeighborsRegressor}): 
\texttt{p} \{1,2\}, with fixed \texttt{n\_neighbors}=1, \texttt{leaf\_size}=8, \texttt{weights}=uniform, \texttt{algorithm}=kd\_tree.

\textbf{Starvation (Classification).} 
a) Random Forest (\texttt{RandomForestClassifier}): 
\texttt{n\_estimators} \{32,128,256\}, 
\texttt{max\_depth} \{None,5,10,20\}, 
\texttt{min\_samples\_split} \{2,5,10,20\}, 
\texttt{criterion} \{gini,entropy,log\_loss\}, 
\texttt{min\_samples\_leaf} \{1,2,5,10,32,128\}, 
\texttt{max\_features} \{None,sqrt,log2\}.  
b) SVM (\texttt{SVC}): 
\texttt{C} \{0.1,1,10,100,1000,10000\}, 
\texttt{kernel} \{linear,poly,rbf,sigmoid\}, 
\texttt{degree} \{2,3,4,5\}, 
\texttt{gamma} \{scale,auto,0.01,0.1,1,10\}, 
\texttt{coef0} \{0,0.1,0.5,1\}.  
c) KNN (\texttt{KNeighborsClassifier}): 
\texttt{p} \{1,2\}, with fixed \texttt{n\_neighbors}=1, \texttt{leaf\_size}=8, \texttt{weights}=uniform, \texttt{algorithm}=kd\_tree.

\section{Derived lightweight tree estimators}\label{appendix: Simplified trees}
\begin{figure*}
    \centering
    \includegraphics[width=0.95\linewidth, trim=0 0 0 0, clip]{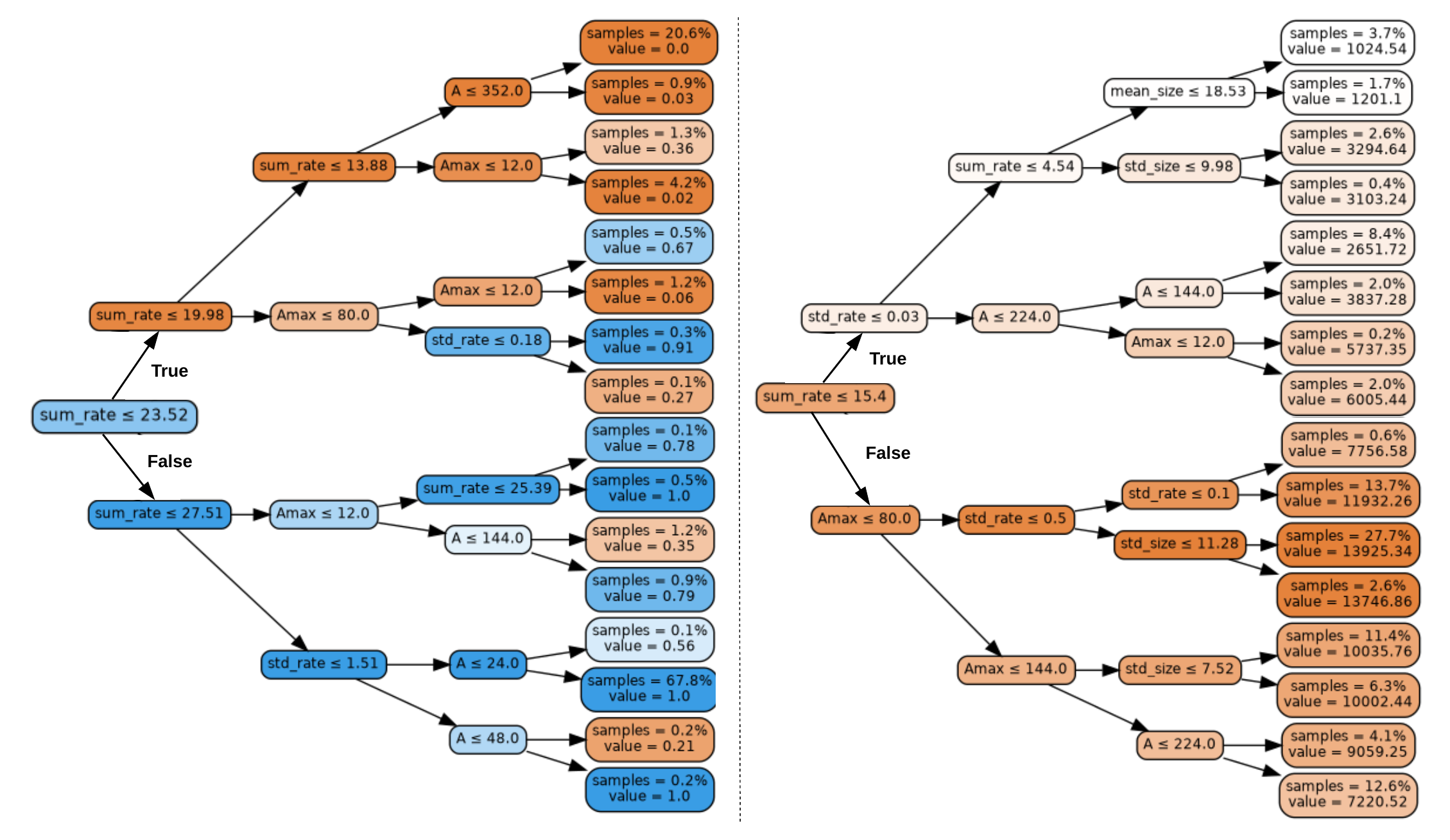}
    \caption{Shallow decision trees derived from the best performing RF model to improve inference performance and enhance model interpretability. (Left) To estimate starvation for Llama-3.1-8B-Instruct. Final value represents the probability of starvation arising. (Right) To estimate throughput for Qwen2.5-7B-Instruct. Final value represents the expected throughput in toks/s.}
    \label{fig:interpretable_trees}
\end{figure*}

Fig.~\ref{fig:interpretable_trees} depicts two of the shallow trees resulting from the refinement phase described in Section~\ref{sec: ML modeling}.

\bibliographystyle{elsarticle-harv}
\bibliography{references}

\end{document}